\documentclass[manuscript]{aastex62}

\usepackage{txfonts}
\usepackage{latexsym,bm}
\usepackage{lineno}
\usepackage{url}
\usepackage{color}
\usepackage{colortbl}
\usepackage{multirow}
\newcommand{\dee}{\mathrm{d}}
\definecolor{mygray}{gray}{.9}

\usepackage[titletoc,title]{appendix}

\newcommand{\qinemail}{qingang@hit.edu.cn}
\newcommand{\hit}{School of Science, Harbin Institute of
	Technology, Shenzhen, 518055, China}
\newcommand{\hitqin}{\hit; \qinemail}

\shorttitle{MOMENTUM DIFFUSION WITH ADIABATIC FOCUSING}
\shortauthors{WANG AND QIN}

\begin{document}
              \arraycolsep 0pt

\title{Study of momentum diffusion with the effect of adiabatic focusing}

\correspondingauthor{G. Qin}
\email{\qinemail}

\author[0000-0002-9586-093X]{J. F. Wang}
\affiliation{\hitqin}

\author[0000-0002-3437-3716]{G. Qin}
\affiliation{\hitqin}

\begin{abstract}
Momentum diffusion of the charged energetic particles
is an important mechanism of the
transport process in astrophysics, 
physics of the fusion devices, and laboratory plasmas.
In addition to the momentum diffusion term for the uniform field,
we obtain an additional  momentum diffusion term
due to the focusing effect of the
large-scale magnetic field.
After evaluating the coefficient of the additional 
momentum diffusion term, 
we find that it is determined by 
the sign of the focusing characteristic length 
and 
the cross helicity of turbulent magnetic field. 
Furthermore, by deriving the mean momentum change rate contributed
from the additional momentum diffusion term, we identify that 
the focused field provides an additional momentum loss or 
gain process.
\end{abstract}


\section{INTRODUCTION}

The transport of the charged energetic particles 
in partially turbulent interstellar 
and interplanetary electromagnetic fields 
has be widely explored in astrophysics, 
the physics of the fusion devices, and gas discharg in 
laboratory plasmas \citep{Parker1965, Jokipii1966,
Icichenko1992,	RuffoloEA1998, ZhangM1999, ZankEA2000, Schlickeiser2002, 
MatthaeusEA2003, 
ZankEA2006, Qin2007, SchlickeiserEA2007, SchlickeiserEA2008, Shalchi2009b, 
SchlickeiserEA2010, Shalchi2010, 
Litvinenko2012a, Litvinenko2012b, 
ZimbardoEA2012, Giacalone2013,
LitvinenkoEA2013, QinEA2014, MalkovEA2015, Malkov2017, Shalchi2017, Shalchi2020}.  
Because of the stochastic property of the particle motion,
the Fokker-Planck equation is used in the description 
of the evolution of the particle distribution function, 
which can incorporate many important effects,
e.g., the pitch-angle scattering, convective process,
the perpendicular diffusion, the adiabatic cooling, 
and the adiabatic focusing as well as the momentum diffusion
\citep{Chandrasekher1943, Jokipii1966, Schlickeiser2002,  SchlickeiserEA2007, 
Shalchi2009b, Schlickeiser2011, LasuikEA2017, Malkov2017, ShalchiEA2019, 
LasuikEA2019}. 

In the light of observations, for the investigation of 
the energetic particle
transport through magnetized plasma,
one usually assumes the
magnetic field configuration as the superposition of the background
magnetic field $B_0$ and the turbulent component $\delta B$.
Consequently, the background field induces a preferred 
direction and leads to the difference of the 
parallel and perpendicular directions
of the background fields 
\citep{Jokipii1966, Schlickeiser2002, Shalchi2009b}. 
If the pitch-angle scattering is strong enough to ensure the length 
characteristic scale of 
the density variation is much greater than the 
mean free path of the charged energetic particles, 
the governing equation of the isotropic distribution function
is usually employed to approximate the Fokker-Planck equation.  
Accordingly, the spatial diffusion coefficients, e.g., 
the parallel diffusion coefficient $\kappa_{zz}$, the perpedicular diffusion
coefficient $\kappa_\perp$, and the drift diffusion coefficient $\kappa_A$, have
been investigated extensively \citep{Jokipii1966, Schlickeiser2002,
MatthaeusEA2003, Shalchi2006, Qin2007, Shalchi2009b, Shalchi2010, TautzEA2012,
Shalchi2020}. 

The momentum diffusion, which expresses the change of the kinetic 
energy of the energetic charged particle distribution function,
is the important process for many scenarios \citep{Parker1965, Kulsrud1979, 
Schlickeiser1989a, Schlickeiser1989b, 
SchlickeiserEA2007, SchlickeiserEA2008, StawarzEA2008, O'SullivanEA2009, 
SchlickeiserEA2010, LefaEA2011, 
MertschEA2011, LeeEA2012, Petrosian2012}.
For the ordered uniform magnetic field $B_0$, 
many scientists have obtained the coefficient of the 
momentum diffusion as  
\citep{Schlickeiser1989a, 
Schlickeiser1989b, Schlickeiser2002, SchlickeiserEA2007, O'SullivanEA2009, 
MertschEA2011}
\begin{eqnarray}
A=\frac{1}{2}\int_{-1}^1d\mu 
\left(D_{pp}-\frac{D_{p\mu}D_{\mu p}}{D_{\mu\mu}}\right),
\label{A} 
\end{eqnarray}
where the quantities $D_{pp}$, $D_{\mu p}$, $D_{\mu \mu}$, and $D_{p\mu}$ are the Fokker-Planck 
coefficients for the homogeneous large-scale magnetic field. 

However, 
the fact that the large-scale magnetic field is nonuniform in space
gives rise to the so called adiabatic focusing effect
of the energetic particles \citep{Parker1958}. Therefore, the focusing Fokker-Planck 
equation should be used to study energetic particles transport
\citep{Roelof1969,
Earl1976, Kunstmann1979, SchlickeiserEA2008, 
SchlickeiserEA2010, Litvinenko2012a, Litvinenko2012b, MalkovEA2015, 
WangEA2016, WangEA2017, wq2018, wq2019, wq2020}. 
The influence of the adiabatic focusing effect to the 
parallel and perpendicular diffusion coefficients have been explored
in the previous papers \citep{BeeckEA1986, BieberEA1990, Kota2000, 
SchlickeiserEA2008, SchlickeiserEA2010, Shalchi2009a, Shalchi2011, 
Litvinenko2012a, Litvinenko2012b, LitvinenkoEA2013, ShalchiEA2013, HeEA2014, 
WangEA2017, wq2018, wq2019, wq2020}.
In addition, some authors \citep{SchlickeiserEA2008, 
SchlickeiserEA2010} found that the along-field focusing 
can leads to an additional convective term in momentum space. 
The mean momentum change rate caused by the additional convective
term was derived \citep{SchlickeiserEA2008, 
SchlickeiserEA2010, LitvinenkoEA2011}
and a new first-order acceleration mechanism
contributed from focused field
was identified \citep{SchlickeiserEA2008, 
SchlickeiserEA2010, LitvinenkoEA2011}.

Furthermore, the focusing effect on momentum diffusion should be investigated. 
Using the perturbation method, 
\citet{SchlickeiserEA2008} investigated
the momentum diffusion for the along-field focusing
and obtained the formula of the momentum diffusion coefficient
same as Equation (\ref{A}), the homogeneous field result, 
which is independent of focusing effect.
Based on this work, \citet{SchlickeiserEA2010}
found similar result to consider perpendicular focusing effect in addition. 
Additionly, the combination effect of 
the  momentum convective term
and the second-order momentum derivative term (SOMT)
for the constant field
was explored and for some special cases
the mean change rate 
of particle momentum was derived
by \citet{ArmstrongEA2012} (ALC2012). 
However, the relationship of the adiabatic focusing effect
to the second-order momentum diffusion term 
was not explored in the paper of ALC2012.   
Therefore, it is an open problem about the contribution of 
the focused field to the SOMT. 

In this paper,  by using the combination of the perturbation method and the iteration technique 
developed previously \citep{WangEA2017, wq2018, wq2019, wq2020},  
we derive the governing equation of the
isotropic distribution function with the adiabatic 
focusing effect. The equation shows that
adiabatic focusing effect gives rise to an additional 
SOMT.
In addition, we obtain
the mean momentum change rate 
contributed from the additional SOMT
and explore the relationship 
between focused field and momentum diffusion.
This paper is organized as follows. In Section 
\ref{The momentum diffusion coefficient for constant background magnetic field}, 
the coefficient of the SOMT for the homogeneous field is deduced. 
In Section \ref{The momentum diffusion coefficient for focusing field}, 
the equation of the isotropic distribution function 
with the adiabatic focusing effect is deduced and the 
coefficient $A(L)$ 
of the additional SOMT
caused by focused field is obtained. 
In Section 
\ref{Evaluating the momentum diffusion coefficient $A$  with the adiabatic focusing effect}, 
the momentum diffusion coefficient $A(L)$
is evaluated.
In Section \ref{Discussion}, 
the mean change rate of particle momentum 
caused by focused field 
is deduced, and 
the physical meaning of the additional SOMT is discussed. 
We conclude
and summarize our results in Section 
\ref{SUMMARY AND CONCLUSION}.

\section{The coefficient of the second-order momentum derivative term for constant background magnetic field}
\label{The momentum diffusion coefficient 
for constant background magnetic field}
In order to obtain the coefficient of the SOMT
for constant field,
we have to first derive the governing equation 
of the isotropic distribution function (EIDF)
for constant field,
which can be obtained from 
the Fokker-Planck equation. 
We start from the Fokker-Planck equation with
the pitch-angle diffusion, the momentum diffusion, and the corresponding cross terms 
\begin{eqnarray}
\frac{\partial{f}}{\partial{t}}+v\mu \frac{\partial{f}}{\partial{z}}=
\frac{\partial{}}{\partial{\mu}}\left(D_{\mu\mu}(\mu)
\frac{\partial{f}}{\partial{\mu}}\right)
+\frac{1}{p^2}\frac{\partial{}}{\partial{\mu}}
\left(p^2D_{\mu p}\frac{\partial{f}}{\partial{p}}\right)
+\frac{1}{p^2}\frac{\partial{}}{\partial{p}}\left(p^2D_{p\mu}
\frac{\partial{f}}{\partial{\mu}}\right)
+\frac{1}{p^2}\frac{\partial{}}{\partial{p}}
\left(p^2D_{pp}\frac{\partial{f}}{\partial{p}}\right),
\label{Schlickeiser2007}
\end{eqnarray}
where $t$ is time, $z$ is the distance along the background
magnetic field, $\mu=v_z /v$ is the pitch-angle
cosine with particle speed $v$ and the  z-component of
the velocity $v_z$, $D_{\mu \mu}(\mu)$ is the
pitch-angle diffusion
coefficient which is assumed to be the function 
of only the pitch-angle cosine  $\mu$,
and $f=f(z,\mu,p,t)$ is the gyrotropic distribution function of energetic particles. 
The terms related to source is ignored
in Equation (\ref{Schlickeiser2007}).
The more complete form of 
the Fokker-Planck equation
can be found in \citet{Schlickeiser2002}.

The distribution function $f=f(z,\mu,p,t)$
can be divided into
its pitch-angle averaged $F(z, p, t)$ and the
anisotropic part $g(z,\mu, p, t)$
\citep[see, e.g.,][]
{SchlickeiserEA2007, SchlickeiserEA2008, HeEA2014,
WangEA2017, wq2018, wq2019, wq2020}
\begin{equation}
f(z, \mu, p, t)=F(z, p, t)+g(z, \mu, p, t)
\label{f=F+g}
\end{equation}
with
\begin{equation}
F(z, p, t)=\frac{1}{2}\int_{-1}^1 d\mu f(z, \mu, p, t)
\label{F and f}
\end{equation}
and
\begin{equation}
\int_{-1}^1 d\mu g(z, \mu, p, t)=0.
\label{integrate g over mu=0}
\end{equation}
Averaging Equation (\ref{Schlickeiser2007})
over $\mu$ from $-1$ to $1$ yields
\begin{eqnarray}
\frac{\partial{F}}{\partial{t}}
+\frac{v}{2}\frac{\partial{}}{\partial{z}}\int_{-1}^1d\mu\mu g
=&&
\frac{1}{p^2}\frac{\partial{}}{\partial{p}}\left[p^2
\frac{1}{2}\int_{-1}^1d\mu \left(D_{pp}\frac{\partial{F}}{\partial{p}}
+D_{p\mu}
\frac{\partial{g}}{\partial{\mu}}
+ D_{pp}\frac{\partial{g}}{\partial{p}}\right)\right],
\label{Schlickeiser2007-averaged by mu}
\end{eqnarray}
where the relation  $D_{\mu\mu}(\mu=\pm 1)=D_{\mu p}(\mu=\pm 1)=0$
is employed \citep{SchlickeiserEA2007, SchlickeiserEA2008, SchlickeiserEA2010}. 
From the latter equation we can obtain 
the EIDF for constant field 
if the anisotropic distribution function
$g(z, \mu, p, t)$ can be found. 

In order to give a concrete example of practical interest, in this paper one
specialize to the isospectral undamped Alfv\'enic slab
turbulence with constant magnetic and cross helicities which
are independent of the wavenumber. 
The cross helicity
$H_C=\left(I^+-I^-\right)/\left(I^++I^-\right)$
indicates the relative intensities
$I^+$ and $I^-$ of the forward- and backward-propagating waves, and
the magnetic helicities
$\sigma^+$ and $\sigma^-$ denotes the polarization states of the forward-
and backward-propagating Alfvén waves. 
With
the quasilinear theory, 
the following formulas are obtained 
\citep{Schlickeiser1989a, DungEA1990, Schlickeiser2002}
\begin{eqnarray}
&&D_{\mu\mu}=\tilde{D}\left(1-\mu^2\right)N,
\label{Dmumu}
\\
&&D_{p\mu}=\epsilon p\tilde{D}\left(1-\mu^2\right)M,
\\
&&D_{\mu p}=\epsilon p\tilde{D}\left(1-\mu^2\right)M,
\\
&&D_{pp}=\epsilon^2 p^2\tilde{D}\left(1-\mu^2\right)R.
\label{Dpp}
\end{eqnarray}
where $\tilde{D}$ is determined by the turbulent spectrum, 
the parameter $\epsilon=v_A/v\ll1$ with the Alv\'enic speed 
$v_A$. The formulas of $N$, $M$, and $R$ are  shown as
\begin{eqnarray}
N(\mu)=&&\left(1+H_C\right)\left(1-\epsilon\mu\right)^2 
|\mu-\epsilon|^{q-1}\left\{\left(1+\sigma^+\right)
S\left[Z\left(\epsilon-\mu\right)\right]
+\left(1-\sigma^+\right)
S\left[Z\left(\mu-\epsilon\right)\right] \right\}
\nonumber\\
&&
+\left(1-H_C\right)\left(1+\epsilon\mu\right)^2 
|\mu+\epsilon|^{q-1}\left\{\left(1+\sigma^-\right)
S\left[-Z\left(\epsilon+\mu\right)\right]
+\left(1-\sigma^+\right)
S\left[Z\left(\mu+\epsilon\right)\right] \right\},
\label{N-helicity}
\end{eqnarray}
\begin{eqnarray}
M(\mu)=&&\left(1+H_C\right)\left(1-\epsilon\mu\right)^2 
|\mu-\epsilon|^{q-1}\left\{\left(1+\sigma^+\right)
S\left[Z\left(\epsilon-\mu\right)\right]
+\left(1-\sigma^+\right)
S\left[Z\left(\mu-\epsilon\right)\right] \right\}
\nonumber\\
&&
-\left(1-H_C\right)\left(1+\epsilon\mu\right)^2 
|\mu+\epsilon|^{q-1}\left\{\left(1+\sigma^-\right)
S\left[-Z\left(\epsilon+\mu\right)\right]
+\left(1-\sigma^+\right)
S\left[Z\left(\mu+\epsilon\right)\right] \right\},
\label{M-helicity}
\end{eqnarray}
\begin{eqnarray}
R(\mu)=&&\left(1+H_C\right)
|\mu-\epsilon|^{q-1}\left\{\left(1+\sigma^+\right)
S\left[Z\left(\epsilon-\mu\right)\right]
+\left(1-\sigma^+\right)
S\left[Z\left(\mu-\epsilon\right)\right] \right\}
\nonumber\\
&&
+\left(1-H_C\right) 
|\mu+\epsilon|^{q-1}\left\{\left(1+\sigma^-\right)
S\left[-Z\left(\epsilon+\mu\right)\right]
+\left(1-\sigma^+\right)
S\left[Z\left(\mu+\epsilon\right)\right] \right\},
\label{R-helicity}
\end{eqnarray}
where $Z$ is the sign of the particle charge, $q$ is the spectral index of the turbulent magnetic field, and 
$S$ is the Heaviside step function.
Here, $N(\mu)$, $M(\mu)$, and $R(\mu)$ have very complicated forms 
and are hard to manipulate. 
To achieve analytical progress, in this paper we 
only assume isotropic pitch-angle scattering and 
take $q=1$. We explore the speed condition
$v_A\ll v\ll c$, i.e., $v/c\ll1$ and $\epsilon=v_A/v\ll1$
and take no net polarization($\sigma^+=\sigma^-=0$).
For these range of the parameters, 
Equations (\ref{Dmumu})-(\ref{Dpp}) becomes
\begin{eqnarray}
&&D_{\mu\mu}=\left(1-\mu^2\right)D_1,
\label{Dmumu-2}\\
&&D_{p\mu}=\epsilon p\left(1-\mu^2\right)D_2,
\label{Dpmu}\\
&&D_{\mu p}=\epsilon p\left(1-\mu^2\right)D_2,
\label{Dmup}\\
&&D_{pp}=\epsilon^2 p^2\left(1-\mu^2\right)D_1,
\label{Dpp-2}
\end{eqnarray}
where $D_1=2\tilde{D}>0$ and $D_2=2\tilde{D}H_C$ 
are determined by the field and 
independent of pitch-angle of particles.
Here, 
\begin{eqnarray}
\tilde{D}=\frac{\pi}{4}\left(s-1\right)vk_{min}
\left(k_{min}R_g\right)^{s-2}\left(\frac{\delta B}{B_0}\right)^2\left(1-\mu^2\right)
\end{eqnarray}
which is a positive real for appropriate parameter range.
Obviously, we can obtain
\begin{eqnarray}
&&\frac{D_2}{D_1}=H_C.
\end{eqnarray}
Because of $-1<H_c<1$, we can find the relation $-1<D_2/D_1<1$ which will be 
used in this paper. 
For simplification we assume the parameters $D_1$
and $D_2$ are all constant \citep{ArmstrongEA2012},
and for convenience in this paper we set the small 
parameter $\epsilon$ as a constant. The case of
the variable $\epsilon$ will be invstigated 
in the future.   

\subsection{The anisotropic distribution function $g(z, \mu, p,t)$}
\label{The anisotropic distribution function}
 
Inserting Equation (\ref{f=F+g}) into Equation (\ref{Schlickeiser2007}) gives
\begin{eqnarray}
\frac{\partial{F}}{\partial{t}}+\frac{\partial{g}}{\partial{t}}
+v\mu \frac{\partial{F}}{\partial{z}}
+v\mu \frac{\partial{g}}{\partial{z}}
=&&
\frac{\partial{}}{\partial{\mu}}\left[D_{\mu\mu}(\mu)
\frac{\partial{g}}{\partial{\mu}}\right]
+\frac{\partial{}}{\partial{\mu}}\left(D_{\mu p}
\frac{\partial{F}}{\partial{p}}\right)
+\frac{\partial{}}{\partial{\mu}}
\left(D_{\mu p}\frac{\partial{g}}{\partial{p}}\right)\nonumber\\
&&+\frac{1}{p^2}\frac{\partial{}}{\partial{p}}\left(p^2D_{p\mu}
\frac{\partial{g}}{\partial{\mu}}\right)
+\frac{1}{p^2}\frac{\partial{}}{\partial{p}}\left(p^2D_{pp}
\frac{\partial{F}}{\partial{p}}\right)
+\frac{1}{p^2}\frac{\partial{}}{\partial{p}}\left(p^2D_{pp}
\frac{\partial{g}}{\partial{p}}\right).
\end{eqnarray}
To rearrange the terms in the latter equation one yield 
\begin{eqnarray}
\frac{\partial{}}{\partial{\mu}}\left[D_{\mu\mu}(\mu)
\frac{\partial{g}}{\partial{\mu}}\right]
=&&\frac{\partial{F}}{\partial{t}}+\frac{\partial{g}}{\partial{t}}
+v\mu \frac{\partial{F}}{\partial{z}}
+v\mu \frac{\partial{g}}{\partial{z}}
-\frac{\partial{}}{\partial{\mu}}\left(D_{\mu p}\frac{\partial{F}}{\partial{p}}\right)
-\frac{\partial{}}{\partial{\mu}}
\left(D_{\mu p}\frac{\partial{g}}{\partial{p}}\right)-\frac{1}{p^2}
\frac{\partial{}}{\partial{p}}\left(p^2D_{p\mu}\frac{\partial{g}}{\partial{\mu}}\right)
\nonumber\\
&&
-\frac{1}{p^2}\frac{\partial{}}{\partial{p}}\left(p^2D_{pp}
\frac{\partial{F}}{\partial{p}}\right)
-\frac{1}{p^2}\frac{\partial{}}{\partial{p}}\left(p^2D_{pp}
\frac{\partial{g}}{\partial{p}}\right).
\label{Schlickeiser2007-rearranged}
\end{eqnarray}
In addition, by integrating Equation (\ref{Schlickeiser2007-rearranged}) over $\mu$
from $-1$ to $\mu$, we obtain 
\begin{eqnarray}
D_{\mu\mu}(\mu)\frac{\partial{g}}{\partial{\mu}}
=&&\frac{\partial{F}}{\partial{t}}\left(\mu+1\right)
+\int_{-1}^{\mu}d\nu\frac{\partial{g}}{\partial{t}}
+v\frac{\mu^2-1}{2} \frac{\partial{F}}{\partial{z}}
+v\int_{-1}^{\mu}d\nu\nu \frac{\partial{g}}{\partial{z}}
-D_{\mu p}\frac{\partial{F}}{\partial{p}}
-D_{\mu p}\frac{\partial{g}}{\partial{p}}
-\frac{1}{p^2}\frac{\partial{}}{\partial{p}}
\left(p^2\int_{-1}^{\mu}d\nu D_{p\nu}\frac{\partial{g}}{\partial{\nu}}\right)
\nonumber\\
&&
-\frac{1}{p^2}\frac{\partial{}}{\partial{p}}
\left(p^2 \int_{-1}^{\mu}d\nu D_{pp}\frac{\partial{F}}{\partial{p}}\right)
-\frac{1}{p^2}\frac{\partial{}}{\partial{p}}
\left(p^2\int_{-1}^{\mu}d\nu D_{pp}\frac{\partial{g}}{\partial{p}}\right).
\label{Schlickeiser2007-average first}
\end{eqnarray}
Here, the regularity $D_{\mu\mu}(\mu=\pm 1)=D_{\mu p}(\mu=\pm 1)=0$
is used \citep{SchlickeiserEA2007,
SchlickeiserEA2008, SchlickeiserEA2010}. Dividing Equation
(\ref{Schlickeiser2007-average first})
by the pitch-angle diffusion coefficient 
$D_{\mu\mu}(\mu)$ yields
\begin{eqnarray}
\frac{\partial{g}}{\partial{\mu}}=\Phi(z,\mu,p,t)
\label{gmu-0}
\end{eqnarray}
with
\begin{eqnarray}
\Phi(z,\mu,p,t)=&&\frac{1}{D_{\mu\mu}}
\Bigg[\frac{\partial{F}}{\partial{t}}\left(\mu+1\right)
+\int_{-1}^{\mu}d\nu\frac{\partial{g}}{\partial{t}}
+v\frac{\mu^2-1}{2} \frac{\partial{F}}{\partial{z}}
+v\int_{-1}^{\mu}d\nu\nu \frac{\partial{g}}{\partial{z}}
-D_{\mu p}\frac{\partial{F}}{\partial{p}}
-D_{\mu p}\frac{\partial{g}}{\partial{p}}
\nonumber\\
&&
-\frac{1}{p^2}\frac{\partial{}}{\partial{p}}
\left(p^2\int_{-1}^{\mu}d\nu D_{p\nu}\frac{\partial{g}}{\partial{\nu}}\right)
-\frac{1}{p^2}\frac{\partial{}}{\partial{p}}
\left(p^2 \int_{-1}^{\mu}d\nu D_{pp}\frac{\partial{F}}{\partial{p}}\right)
-\frac{1}{p^2}\frac{\partial{}}{\partial{p}}
\left(p^2\int_{-1}^{\mu}d\nu D_{pp}\frac{\partial{g}}{\partial{p}}\right)\Bigg].
\label{Phi-for constant field}
\nonumber\\
\end{eqnarray}
By integrating Equation (\ref{gmu-0}) over $\mu$ from $-1$ to $\mu$, one 
can find
\begin{eqnarray}
g(z,\mu,p,t)=g(-1)+\int_{-1}^{\mu}d\nu \Phi(z,\nu,p,t).
\label{g-2 for constant field}
\end{eqnarray}
To proceed, by manipulating the integration 
$1/2\int_{-1}^1d\mu$ to the latter equation, we can obtain
\begin{eqnarray}
&&0=2g(-1)+\int_{-1}^1d\mu\int_{-1}^{\mu}d\nu \Phi(z,\nu,p,t). 
\end{eqnarray}
Therefore, we can find
\begin{eqnarray}
&&g(-1)=-\frac{1}{2}\int_{-1}^1d\mu\int_{-1}^{\mu}d\nu 
\Phi(z,\nu,p,t)
=-\frac{1}{2}\int_{-1}^{1}d\mu (1-\mu)
\Phi(z,\mu,p,t).
\end{eqnarray}
Inserting the latter formula 
into Equation (\ref{g-2 for constant field}) gives
\begin{eqnarray}
&&g(z,\mu,p,t)=
\int_{-1}^{\mu}d\mu \Phi(z,\mu,p,t)
-\frac{1}{2}\int_{-1}^{1}d\mu (1-\mu)\Phi(z,\mu,p,t),
\label{g-constant field-D is constant}
\end{eqnarray}
which is the formula of 
the anisotropic distribution function
for constant background magnetic field.

\subsection{The isotropic distribution function equation for constant field}
\label{The isotropic distribution function equation for constant field}

By combininging Equations
(\ref{Schlickeiser2007-averaged by mu}),
(\ref{Phi-for constant field}) 
and 
(\ref{g-constant field-D is constant}), 
we can easily obtain
\begin{eqnarray}
\frac{\partial{F}}{\partial{t}}
&&=\frac{1}{p^2}\frac{\partial{}}{\partial{p}}
\left(p^2
A
\frac{\partial{F}}{\partial{p}}\right)
+\frac{1}{p^2}\frac{\partial{}}{\partial{p}}
\left(p^2
A'
\frac{\partial{F}}{\partial{p}}\right)
+\frac{1}{p^2}\frac{\partial{}}{\partial{p}}
\Bigg[p^2\kappa_{3p1}^c
\frac{\partial{}}{\partial{p}}
\left(\kappa_{3p2}^c
\frac{\partial{F}}{\partial{p}}\right)\Bigg]
+\frac{1}{p^2}\frac{\partial{}}{\partial{p}}
\Bigg\{p^2
\kappa_{4p1}^c\frac{\partial{}}{\partial{p}}
\Bigg[\kappa_{4p2}^c\frac{\partial{}}{\partial{p}}
\left(\kappa_{4p3}^c\frac{\partial{F}}{\partial{p}}
\right)\Bigg]\Bigg\}
+\cdots,
\label{equation of F for constant field}
\end{eqnarray}
Here, the parameters $A_0$, 
$\kappa_{3p1}^c$, $\kappa_{3p2}^c$, $\cdots$,
are the coefficients of momentum derivative terms
for constant field,
and the coefficient $A_0$ 
of the SOMT 
is the more general form of  
the coefficient $A$
(see Equation (\ref{A})).
Obviously, Equation 
(\ref{equation of F for constant field})
does not contain 
the first-order momentum 
derivative term.

\subsection{The momentum diffusion coefficient for the lowest order of $\epsilon$}
\label{The lowest order momentum diffusion coefficient}

Because the momentum diffusion term only contain the derivative 
with respect to  momentum $p$, 
we can find that 
only the terms on 
the right hand side of Equation 
(\ref{Schlickeiser2007-averaged by mu})
\begin{eqnarray}
\frac{1}{p^2}\frac{\partial{}}{\partial{p}}\left[p^2
\frac{1}{2}\int_{-1}^1d\mu \left(D_{pp}\textcircled{2}\frac{\partial{F}}{\partial{p}}
+D_{p\mu}\textcircled{1}
\frac{\partial{g}}{\partial{\mu}}
+ D_{pp}\textcircled{2}\frac{\partial{g}}{\partial{p}}\right)\right]
\label{right hand side to A}
\end{eqnarray}
can contribute to the SOMT.
Here, for the convenience of discussion,
we mark the lowest order of small quantity $\epsilon$
by the circled number. For example,
$D_{pp}\textcircled{2}$ indicates
the lowest order of $D_{pp}$ is $\epsilon^2$,
and $D_{p\mu}\textcircled{1}$ denotes 
the lowest order of $D_{p\mu}$ is $\epsilon^1$. 
This symbol is very helpful in the following deduction and used 
in the entire paper. 
In order to derive the coefficient of the SOMT
from the above expression,
the terms in $\partial{g}/\partial{\mu}$ and $\partial{g}/\partial{p}$
have to to found.

Using Equation (\ref{g-constant field-D is constant}),
we can obtain the following formula
\begin{eqnarray}
&&\frac{\partial{g}}{\partial{\mu}}=\frac{\partial{}}{\partial{\mu}}
\Bigg[\int_{-1}^{\mu}d\nu \Phi(z,\nu,p,t)
-\frac{1}{2}\int_{-1}^{1}d\mu (1-\mu)\Phi(z,\mu,p,t)\Bigg]
=\Phi(z,\mu,p,t).
\label{g/mu-constant field-constan D-0}
\end{eqnarray}
Combining Equations (\ref{Phi-for constant field}) 
and (\ref{g/mu-constant field-constan D-0}) gives
\begin{eqnarray}
\frac{\partial{g}}{\partial{\mu}}
=&&\frac{1}{D_{\mu\mu}}\Bigg[\frac{\partial{F}}{\partial{t}}\left(\mu+1\right)
+\int_{-1}^{\mu}d\nu\frac{\partial{g}}{\partial{t}}
+v\frac{\mu^2-1}{2} \frac{\partial{F}}{\partial{z}}
+v\int_{-1}^{\mu}d\nu\nu \frac{\partial{g}}{\partial{z}}
-D_{\mu p}\frac{\partial{F}}{\partial{p}}
-D_{\mu p}\frac{\partial{g}}{\partial{p}}
\nonumber\\
&&
-\frac{1}{p^2}\frac{\partial{}}{\partial{p}}
\left(p^2\int_{-1}^{\mu}d\nu D_{p\nu}\frac{\partial{g}}{\partial{\nu}}\right)
-\frac{1}{p^2}\frac{\partial{}}{\partial{p}}
\left(p^2 \int_{-1}^{\mu}d\nu D_{pp}\frac{\partial{F}}{\partial{p}}\right)
-\frac{1}{p^2}\frac{\partial{}}{\partial{p}}
\left(p^2\int_{-1}^{\mu}d\nu D_{pp}\frac{\partial{g}}{\partial{p}}\right)\Bigg].
\label{g/mu-100}
\end{eqnarray}
Because the first and second terms 
on the right-hand side of the latter equation contain
the operator $\partial{}/\partial{t}$,
and the third and fourth terms 
have the operator
$\partial{}/\partial{z}$,
they cannot 
contribute to the term of $\partial{F}/\partial{p}$. 
For the sake of simplicity, in 
the following deduction we ignore the
terms not contributing to  
$\partial{F}/\partial{p}$, that is,
neglect the terms containing $\partial{}/\partial{t}$
and $\partial{}/\partial{z}$.
Accordingly, we use ``$\Longrightarrow$"
to relapce the equal sign ``$=$" 
in the corresponding equations. 
Thus, Equation (\ref{g/mu-100}) becomes
\begin{eqnarray}
\frac{\partial{g}}{\partial{\mu}}&&\Longrightarrow 
-\frac{D_{\mu p}}{D_{\mu\mu}}\textcircled{1}
\frac{\partial{F}}{\partial{p}}
-\frac{D_{\mu p}}{D_{\mu\mu}}\textcircled{1}
\frac{\partial{g}}{\partial{p}}
-\frac{1}{D_{\mu\mu}}\frac{1}{p^2}\frac{\partial{}}{\partial{p}}
\left(p^2\int_{-1}^{\mu}d\nu D_{p\nu}\textcircled{1}
\frac{\partial{g}}{\partial{\nu}}\right)
-\frac{1}{D_{\mu\mu}}\frac{1}{p^2}\frac{\partial{}}{\partial{p}}
\left(p^2 \int_{-1}^{\mu}d\nu D_{pp}\textcircled{2}\frac{\partial{F}}{\partial{p}}\right)
\nonumber\\
&&-\frac{1}{D_{\mu\mu}}\frac{1}{p^2}\frac{\partial{}}{\partial{p}}
\left(p^2\int_{-1}^{\mu}d\nu D_{pp}\textcircled{2}
\frac{\partial{g}}{\partial{p}}\right).
\label{g/mu-constant field-constan D-1}
\end{eqnarray}
If inserting the latter equation into 
formula (\ref{right hand side to A}), we can find that
the lowest order is $\epsilon^2$. 
For simplification, in this subsection  
we only deduce the coefficient of the SOMT exact up to
second order, i.e., $\epsilon^2$.  
So, the formula of 
$\partial{g}/\partial{\mu}$ can be exact up to 
$\epsilon^1$ and the higher-order terms in 
$\partial{g}/\partial{\mu}$ are ignored. 
Thus, formula 
(\ref{g/mu-constant field-constan D-1}) becomes  
\begin{eqnarray}
\frac{\partial{g}}{\partial{\mu}}&&\Longrightarrow 
-\frac{D_{\mu p}}{D_{\mu\mu}}\textcircled{1}
\frac{\partial{F}}{\partial{p}}
-\frac{D_{\mu p}}{D_{\mu\mu}}\textcircled{1}
\frac{\partial{g}}{\partial{p}}
-\frac{1}{D_{\mu\mu}}\frac{1}{p^2}\frac{\partial{}}{\partial{p}}
\left(p^2\int_{-1}^{\mu}d\nu
D_{p\nu}\textcircled{1}\frac{\partial{g}}{\partial{\nu}}\right),
\label{g/mu-constant field-constan D-2}
\end{eqnarray}
which contains $\partial{g}/\partial{p}$
and $\partial{g}/\partial{\mu}$ 
in the third term.
Considering Equations (\ref{Phi-for constant field})
and (\ref{g-constant field-D is constant}),
we can find
\begin{eqnarray}
\frac{\partial{g}}{\partial{p}}
&&=\int_{-1}^{\mu}d\nu \frac{\partial{}}{\partial{p}}\Phi(\nu, p)
-\frac{1}{2}\int_{-1}^{1}d\mu (1-\mu)\frac{\partial{}}{\partial{p}}\Phi(\mu, p)
\label{g/p-2}
\end{eqnarray}
with
\begin{eqnarray}
\frac{\partial{}}{\partial{p}}\Phi(\mu, p)
&&\Longrightarrow
-D_{\mu p}\textcircled{1}\frac{\partial{F}}{\partial{p}}
-D_{\mu p}\textcircled{1}\frac{\partial{g}}{\partial{p}}
-\frac{1}{p^2}\frac{\partial{}}{\partial{p}}
\left(p^2\int_{-1}^{\mu}d\nu
D_{p\nu}\textcircled{1}\frac{\partial{g}}{\partial{\nu}}\right)
-\frac{1}{p^2}\frac{\partial{}}{\partial{p}}
\left(p^2 \int_{-1}^{\mu}d\nu D_{pp}\textcircled{2}\frac{\partial{F}}{\partial{p}}\right)
\nonumber\\
&&
-\frac{1}{p^2}\frac{\partial{}}{\partial{p}}
\left(p^2\int_{-1}^{\mu}d\nu D_{pp}\textcircled{2}\frac{\partial{g}}{\partial{p}}\right)
\label{phi/p}.
\end{eqnarray}
The latter formula shows that the lowest order 
of $\partial{g}/\partial{p}$ is the first order 
$\epsilon^1$. 
So, Equation 
(\ref{g/mu-constant field-constan D-2}) becomes
\begin{eqnarray}
\frac{\partial{g}}{\partial{\mu}}&&\Longrightarrow
-\frac{D_{\mu p}}{D_{\mu\mu}}\textcircled{1}
\frac{\partial{F}}{\partial{p}}
-\epsilon^2 \textcircled{2} \frac{\partial{F}}{\partial{p}}
-\frac{1}{D_{\mu\mu}}\frac{1}{p^2}\frac{\partial{}}{\partial{p}}
\left(p^2\int_{-1}^{\mu}d\nu
D_{p\nu}\textcircled{1}\frac{\partial{g}}{\partial{\nu}}\right).
\end{eqnarray}
Because only retaining $\partial{g}/\partial{\mu}$ 
exact up to first order $\epsilon^1$ in this 
subsection,
we can ignore the second term on the right hand 
side of the latter equation. 
Thus, we can obtain 
\begin{eqnarray}
\frac{\partial{g}}{\partial{\mu}}\Longrightarrow 
-\frac{D_{\mu p}}{D_{\mu\mu}}\textcircled{1}
\frac{\partial{F}}{\partial{p}}
-\frac{1}{D_{\mu\mu}}\frac{1}{p^2}\frac{\partial{}}{\partial{p}}
\left(p^2\int_{-1}^{\mu}d\nu D_{p\nu}\textcircled{1}
\frac{\partial{g}}{\partial{\nu}}\right).
\label{g/mu-constant field-g/mu-10}
\end{eqnarray}
The integrand of the second term on the right hand side of the
latter formula contains $\partial{g}/\partial{\mu}$.
If iterating formula (\ref{g/mu-constant field-g/mu-10})
into its right hand side of it, 
we can find
\begin{eqnarray}
\frac{\partial{g}}{\partial{\mu}}\Longrightarrow 
-\frac{D_{\mu p}}{D_{\mu\mu}}\textcircled{1}
\frac{\partial{F}}{\partial{p}}
+\frac{1}{D_{\mu\mu}}\frac{1}{p^2}\frac{\partial{}}{\partial{p}}
\Bigg\{p^2\int_{-1}^{\mu}d\nu D_{p\nu}\textcircled{1}
\Bigg[\frac{D_{\mu p}}{D_{\mu\mu}}\textcircled{1}
\frac{\partial{F}}{\partial{p}}
+\frac{1}{D_{\mu\mu}}\frac{1}{p^2}\frac{\partial{}}{\partial{p}}
\left(p^2\int_{-1}^{\mu}d\nu D_{p\nu}\textcircled{1}
\frac{\partial{g}}{\partial{\nu}}\right)\Bigg]\Bigg\}.
\label{g/mu-constant field-g/mu-11}
\end{eqnarray}
Obviously, in order to retain
$\partial{g}/\partial{\mu}$
exact up to first order $\epsilon^1$,
formula (\ref{g/mu-constant field-g/mu-11}) becomes
\begin{eqnarray}
\frac{\partial{g}}{\partial{\mu}}\Longrightarrow 
-\frac{D_{\mu p}}{D_{\mu\mu}}\textcircled{1}
\frac{\partial{F}}{\partial{p}}.
\label{g/mu-constant field-g/mu-12}
\end{eqnarray}

Inserting formulas (\ref{g/p-2}) and 
(\ref{g/mu-constant field-g/mu-12}) into 
the right hand side of Equation 
(\ref{Schlickeiser2007-averaged by mu}) yields
\begin{eqnarray}
\frac{\partial{F}}{\partial{t}}
+\frac{v}{2}\frac{\partial{}}{\partial{z}}\int_{-1}^1d\mu\mu g
\sim&&
\frac{1}{p^2}\frac{\partial{}}{\partial{p}}
\Bigg[p^2
\frac{1}{2}\int_{-1}^1d\mu \left(D_{pp}\textcircled{2}
-
\frac{D_{\mu p}D_{p\mu}}{D_{\mu\mu}}\textcircled{2}
+ D_{pp}\epsilon\textcircled{3} \right)\Bigg]
\frac{\partial{F}}{\partial{p}}. 
\end{eqnarray}
Accordingly, we can obtain the coefficient of the SOMT exact up to $\epsilon^2$ as
\begin{eqnarray}
A=\frac{1}{2}\int_{-1}^1d\mu 
\left(D_{pp}-\frac{D_{p\mu}D_{\mu p}}{D_{\mu\mu}}\right),
\label{A by us} 
\end{eqnarray}
which is identical with the result derived 
by the previous
researchers \citep{Schlickeiser2002, SchlickeiserEA2007, SchlickeiserEA2008, 
SchlickeiserEA2010}. Of course, 
Equation (\ref{A by us}) is also the lowest order form of 
the SOMT coefficient.

\section{The momentum diffusion coefficient for focusing field}
\label{The momentum diffusion coefficient for focusing field}

The spatially varying mean magnetic field
gives rise to the so-called particle adiabatic focusing process
and influences the spatial parallel and perpendicular diffusion
\citep{BeeckEA1986, BieberEA1990, Kota2000, 
SchlickeiserEA2008, SchlickeiserEA2010, Shalchi2009a, Shalchi2011, 
Litvinenko2012a, Litvinenko2012b, LitvinenkoEA2013, ShalchiEA2013, HeEA2014, 
WangEA2017, wq2018, wq2019, wq2020}.
In this paper, we start from the modified Fokker-Planck equation
\begin{eqnarray}
\frac{\partial{f}}{\partial{t}}+v\mu \frac{\partial{f}}{\partial{z}}=
\frac{\partial{}}{\partial{\mu}}\left[D_{\mu\mu}(\mu)
\frac{\partial{f}}{\partial{\mu}}\right]
-\frac{v}{2L}\frac{\partial{}}{\partial{\mu}}\left[\left(1-\mu^2\right)f\right]
+\frac{1}{p^2}\frac{\partial{}}{\partial{\mu}}
\left(p^2D_{\mu p}\frac{\partial{f}}{\partial{p}}\right)
+\frac{1}{p^2}\frac{\partial{}}{\partial{p}}\left(p^2D_{p\mu}
\frac{\partial{f}}{\partial{\mu}}\right)
+\frac{1}{p^2}\frac{\partial{}}{\partial{p}}\left(p^2D_{pp}
\frac{\partial{f}}{\partial{p}}\right)\nonumber\\
\label{modified Fokker-Planck equation-0}
\end{eqnarray}
with the pitch-angle diffusion term, the momentum 
diffusion one, the cross terms and the term with
the adiabatic focusing effect.

Averaging Equation (\ref{modified Fokker-Planck equation-0})
over pitch-angle cosine $\mu$ from $-1$ to $1$ yields
\begin{eqnarray}
&&\frac{\partial{F}}{\partial{t}}+\frac{v}{2}
\frac{\partial{}}{\partial{z}}\int_{-1}^1d\mu\mu g
=\frac{1}{p^2}\frac{\partial{}}{\partial{p}}\left[p^2
\frac{1}{2}\int_{-1}^1d\mu \left(D_{pp}\frac{\partial{F}}{\partial{p}}
+D_{p\mu}
\frac{\partial{g}}{\partial{\mu}}
+ D_{pp}\frac{\partial{g}}{\partial{p}}\right)\right].
\label{continuity equation-constant D for focusing field}
\end{eqnarray}
In order to obtain the coefficient of the SOMT, 
we have to first derive the SOMT which can be written as
\begin{eqnarray}
T_{pp}(L)=\frac{1}{p^2}\frac{\partial{}}{\partial{p}}\left(p^2A(L)
\frac{\partial{F}}{\partial{p}}\right),
\label{A-L}
\end{eqnarray}
where $A(L)$ is the coefficient of the SOMT. 
It is obvious that 
$T_{pp}(L)$ is only contributed from 
the three terms on the right hand side of Equation 
(\ref{continuity equation-constant D for focusing field}).
So, the formulas of 
$\partial{g}/\partial{\mu}$ and $\partial{g}/\partial{p}$
need to be derived. Therefore, the anisotropic distribution 
function $g(z,\mu,p,t)$ with the adiabatic focusing effect 
has to be deduced first.  

\subsection{The anisotropic distribution function $g(z,\mu,p,t)$
with the focusing effect}
\label{The anisotropic distribution function with focusing 
effect}

Here, we employ the iteration method developed in the previous papers
\citep{WangEA2017, wq2018, wq2019, wq2020} to
derive the anisotropic distribution function 
$g(z,\mu,p,t)$ with the focusing effect. 

Combining Equations (\ref{f=F+g}) and 
(\ref{modified Fokker-Planck equation-0})
gives
\begin{eqnarray}
\frac{\partial{F}}{\partial{t}}+\frac{\partial{g}}{\partial{t}}
+v\mu \frac{\partial{F}}{\partial{z}}+v\mu \frac{\partial{g}}{\partial{z}}=&&
\frac{\partial{}}{\partial{\mu}}\left[D_{\mu\mu}(\mu)\frac{\partial{g}}{\partial{\mu}}
-\frac{v}{2L}\left(1-\mu^2\right)F-\frac{v}{2L}\left(1-\mu^2\right)g
+D_{\mu p}\frac{\partial{F}}{\partial{p}}
+D_{\mu p}\frac{\partial{g}}{\partial{p}}\right]\nonumber\\
&&+\frac{1}{p^2}\frac{\partial{}}{\partial{p}}p^2\left(D_{p\mu}
\frac{\partial{g}}{\partial{\mu}}
+D_{pp}\frac{\partial{F}}{\partial{p}}
+D_{pp}\frac{\partial{g}}{\partial{p}}\right).
\label{modified Fokker-Planck equation-f=F+g}
\end{eqnarray}
By manipulating the integration over $\mu$ from 
$-1$ to $\mu$, we can obtain 
\begin{eqnarray}
\frac{\partial{F}}{\partial{t}}\left(\mu+1\right)&&+\frac{\partial{}}{\partial{t}}
\int_{-1}^{\mu}d\nu g
+v\frac{\mu^2-1}{2} \frac{\partial{F}}{\partial{z}}
+v\frac{\partial{}}{\partial{z}}\int_{-1}^{\mu}d\nu\nu g =
D_{\mu\mu}(\mu)\frac{\partial{g}}{\partial{\mu}}
-\frac{v}{2L}\left(1-\mu^2\right)F-\frac{v}{2L}\left(1-\mu^2\right)g
\nonumber\\
&&+D_{\mu p}\frac{\partial{F}}{\partial{p}}
+D_{\mu p}\frac{\partial{g}}{\partial{p}}
+\frac{1}{p^2}\frac{\partial{}}{\partial{p}}p^2
\int_{-1}^\mu d\nu\left(D_{p\nu}\frac{\partial{g}}{\partial{\nu}}
+D_{pp}\frac{\partial{F}}{\partial{p}}
+D_{pp}\frac{\partial{g}}{\partial{p}}\right).
\label{int -1 mu}
\end{eqnarray}
For $\mu=1$, the latter equation becomes
\begin{eqnarray}
\frac{\partial{F}}{\partial{t}}+\frac{v}{2}
\frac{\partial{}}{\partial{z}}\int_{-1}^1d\mu\mu g=
\frac{1}{p^2}\frac{\partial{}}{\partial{p}}\left[p^2
\frac{1}{2}\int_{-1}^1d\mu
\left(D_{p\mu}\frac{\partial{g}}{\partial{\mu}}
+D_{pp}\frac{\partial{F}}{\partial{p}}
+D_{pp}\frac{\partial{g}}{\partial{p}}\right)\right].
\label{continuity equation-1}
\end{eqnarray}
To substract Equation (\ref{continuity equation-1})
from Equation (\ref{int -1 mu}) we find
\begin{eqnarray}
\frac{\partial{F}}{\partial{t}}\mu&&+\frac{\partial{}}{\partial{t}}\int_{-1}^{\mu}d\nu g
+v\frac{\mu^2-1}{2} \frac{\partial{F}}{\partial{z}}
+v\frac{\partial{}}{\partial{z}}\int_{-1}^{\mu}d\nu\nu g -\frac{v}{2}
\frac{\partial{}}{\partial{z}}\int_{-1}^1d\mu\mu g=
D_{\mu\mu}(\mu)\frac{\partial{g}}{\partial{\mu}}
-\frac{v}{2L}\left(1-\mu^2\right)F-\frac{v}{2L}\left(1-\mu^2\right)g
\nonumber\\
&&+D_{\mu p}\frac{\partial{F}}{\partial{p}}
+D_{\mu p}\frac{\partial{g}}{\partial{p}}
+\frac{1}{p^2}\frac{\partial{}}{\partial{p}}p^2
\int_{-1}^\mu d\nu\left(D_{p\nu}\frac{\partial{g}}{\partial{\nu}}
+D_{pp}\frac{\partial{F}}{\partial{p}}
+D_{pp}\frac{\partial{g}}{\partial{p}}\right)
-\frac{1}{p^2}\frac{\partial{}}{\partial{p}}p^2
\frac{1}{2}\int_{-1}^1d\mu
\left(D_{p\mu}\frac{\partial{f}}{\partial{\mu}}
+D_{pp}\frac{\partial{f}}{\partial{p}}\right).\nonumber\\
\label{int -1 mu - int -1 1}
\end{eqnarray}
{\bf
Equation (\ref{int -1 mu - int -1 1}) can be rewritten as} 
\begin{eqnarray}
\frac{\partial{}}{\partial{\mu}}
\Bigg\{\left[g(z,\mu,p,t)-L\left(\frac{\partial{F}}{\partial{z}}
-\frac{F}{L}\right) \right]e^{-M(\mu)}\Bigg\}
=e^{-M(\mu)}\Phi(z,\mu,p,t),
\label{HS2014}
\end{eqnarray}
{\bf
which was first obtained by \citet{HeEA2014}. 
}
Here, the formulas of $\Phi(z,\mu,p,t)$ and $M(\mu)$ are shown as 
\begin{eqnarray}
\Phi(z,\mu,p,t)&&=\frac{1}{D_{\mu\mu}(\mu)}
\Bigg[\frac{\partial{F}}{\partial{t}}\mu+\frac{\partial{}}{\partial{t}}
\int_{-1}^{\mu}d\nu g
+v\frac{\partial{}}{\partial{z}}\int_{-1}^{\mu}d\nu\nu g -\frac{v}{2}
\frac{\partial{}}{\partial{z}}\int_{-1}^1d\mu\mu g
-D_{\mu p}\frac{\partial{F}}{\partial{p}}
-D_{\mu p}\frac{\partial{g}}{\partial{p}}
\nonumber\\
&&
-\frac{1}{p^2}\frac{\partial{}}{\partial{p}}p^2
\int_{-1}^\mu d\nu\left(D_{p\nu}\frac{\partial{g}}{\partial{\nu}}
+D_{pp}\frac{\partial{F}}{\partial{p}}
+D_{pp}\frac{\partial{g}}{\partial{p}}\right)
+\frac{1}{p^2}\frac{\partial{}}{\partial{p}}p^2
\frac{1}{2}\int_{-1}^1d\mu
\left(D_{pp}\frac{\partial{F}}{\partial{p}}
+D_{p\mu}\frac{\partial{g}}{\partial{\mu}}
+D_{pp}\frac{\partial{g}}{\partial{p}}
\right)\Bigg]
\nonumber\\
\label{Phi}
\end{eqnarray}
and 
\begin{eqnarray}
M(\mu)=\frac{v}{2L}\int_{-1}^{\mu}d\nu
\frac{1-\nu^2}{D_{\nu\nu}(\nu)}.
\label{M}
\end{eqnarray}

Integrating Equation (\ref{HS2014}) 
with respect to $\mu$ from $-1$ to $\mu$
gives
\begin{eqnarray}
&&
g(z,\mu,p,t)
-L\left(\frac{\partial{F}}{\partial{z}}-\frac{F}{L}\right)
-\left[g(\mu=-1)-L\left(\frac{\partial{F}}{\partial{z}}-\frac{F}{L}\right)\right]
e^{M(\mu)}
=e^{M(\mu)}R(z,\mu,p,t) 
\label{2}
\end{eqnarray}
with
\begin{eqnarray}
R(z,\mu,p,t)=\int_{-1}^\mu d\nu e^{-M(\nu)} \Phi(z,\nu,p,t).
\label{R}
\end{eqnarray}
By operating the integration  $\int_{-1}^1d\mu$
over Equation (\ref{2}), we obtain
\begin{eqnarray}
g(\mu=-1)=L\left(\frac{\partial{F}}{\partial{z}}-\frac{F}{L}\right)
-2L\left(\frac{\partial{F}}{\partial{z}}-\frac{F}{L}\right)
\frac{1}{\int_{-1}^1d\mu e^{M(\mu)}}
-\frac{1}{\int_{-1}^1d\mu e^{M(\mu)}}\int_{-1}^1d\mu e^{M(\mu)}R(z,\mu,p,t),
\end{eqnarray}
where Equation (\ref{integrate g over mu=0}) is used.
To insert the latter equation into Equation (\ref{2}), we can obtain
\begin{eqnarray}
g(z,\mu,p,t)=L\left(\frac{\partial{F}}
{\partial{z}}
-\frac{F}{L}\right)\left[1-
\frac{2e^{M(\mu)}}{\int_{-1}^{1}
d\mu e^{M(\mu) }}\right]
+e^{M(\mu)}\left[R(z,\mu,p,t)
-\frac{\int_{-1}^{1}d\mu
e^{M(\mu)}R(z,\mu,p,t)}
{\int_{-1}^{1}d\mu e^{M(\mu) }}\right].
\label{g-focusing field}
\end{eqnarray}

\subsection{The isotropic distribution function equation 
for focused field}
\label{The isotropic distribution function equation 
for focused field}

By using the same operation in Subsection 
\ref{The isotropic distribution function equation for constant field} and employing Equation 
(\ref{g-focusing field}), we can obtain 
the isotropic distribution function
equation for focused field as
\begin{eqnarray}
\frac{\partial{F}}{\partial{t}}
=&&\frac{1}{p^2}\frac{\partial{}}{\partial{p}}
\Bigg(p^2\kappa_p^f F\Bigg)
+\frac{1}{p^2}\frac{\partial{}}{\partial{p}}
\left(p^2
A
\frac{\partial{F}}{\partial{p}}\right)
+\frac{1}{p^2}\frac{\partial{}}{\partial{p}}
\left(p^2
A'
\frac{\partial{F}}{\partial{p}}\right)
+\frac{1}{p^2}\frac{\partial{}}{\partial{p}}
\left(p^2
\mathcal{M}_4(\epsilon,\xi)
\frac{\partial{F}}{\partial{p}}\right)
+\frac{1}{p^2}\frac{\partial{}}{\partial{p}}
\Bigg[p^2\kappa_{3p1}^f
\frac{\partial{}}{\partial{p}}
\left(\kappa_{3p2}^f
\frac{\partial{F}}{\partial{p}}\right)\Bigg]
\nonumber\\
&&
+\frac{1}{p^2}\frac{\partial{}}{\partial{p}}
\Bigg\{p^2
\kappa_{4p1}^f\frac{\partial{}}{\partial{p}}
\Bigg[\kappa_{4p2}^f\frac{\partial{}}{\partial{p}}
\left(\kappa_{4p3}^f\frac{\partial{F}}{\partial{p}}
\right)\Bigg]\Bigg\}
+\cdots,
\label{equation of F for focused field}
\end{eqnarray}
where $\kappa_p^f$, $\mathcal{M}_4(\epsilon,\xi)$,
$\kappa_{3p1}^f$, $\kappa_{3p2}^f$, $\cdots$,
are the coefficients of the momentum derivative terms
for focused field. 
It is obvious that 
focused field contributes to an additional momentum
streaming term which was found by 
\citet{SchlickeiserEA2008} and 
\citet{LitvinenkoEA2011}.
In the following part of this paper, we will explore 
the influence of focused field on the SOMT. 

\subsection{Dimensional analysis of the 
Fokker-Planck equation with focusing effect}
\label{Accuracy analysis of the modifying factors
to the momentum diffusion coefficient}

Here, we set $F\sim O(1)$ and $g\sim \epsilon\ll 1$ because of $F\gg g$.
In addition, we assume 
$z=z'Z$, $t=t'T$, and $p=p'P$ 
with the characteristic scales 
$Z\sim F/|\partial{F}/\partial{z}|$,
$T\sim F/|\partial{F}/\partial{t}|$,
and  
$P\sim F/|\partial{F}/\partial{p}|$,
where $z'$, $t'$, $p'$ are the 
dimensionless quantities.
We also suppose  that the relation $p/P\sim O(1)$ holds. 
The mean free path of the particles is presented as
$\lambda=v\tau$ 
with particle speed $v$ and the characteristic time $\tau$ of the interaction 
of particle and
turbulent magnetic field, i.e., $\tau\sim 1/D_{\mu\mu}$. 
For kinetic description, we assume 
$\lambda/Z=\epsilon\ll 1$.
Furthermore, the ratio of the mean free path $\lambda$ and the adiabatic 
focusing characteristic length $L$,
i.e., the focusing parameter $\xi=\lambda/L$, is assumed as
$\xi\sim\epsilon\ll1$.
In order to satisfy the scale analysis requirements,
we have to set
$\tau/T\sim 
\epsilon^2$. 
Using the above nondimensionaling regulations, we can rewrite 
Equation (\ref{continuity equation-constant D for focusing field}) 
as
\begin{eqnarray}
\frac{\partial{F}}{\partial{t'}}
+\frac{Tv}{2Z}\frac{\partial{}}{\partial{z'}}\int_{-1}^1d\mu\mu g
\sim&&
\frac{T}{p^2}\frac{1}{P}\frac{\partial{}}{\partial{p'}}\left(p^2
\frac{1}{2}\int_{-1}^1d\mu \frac{\epsilon^2
p^2}{\tau}\frac{1}{P}\frac{\partial{F}}{\partial{p'}}\right)
+\frac{T}{p^2}\frac{1}{P}\frac{\partial{}}{\partial{p'}}\left(p^2
\frac{1}{2}\int_{-1}^1d\mu \frac{\epsilon p}{\tau}
\frac{\partial{g}}{\partial{\mu}}\right)
\nonumber\\
&&+\frac{T}{p^2}\frac{1}{P}\frac{\partial{}}{\partial{p'}}\left(p^2
\frac{1}{2}\int_{-1}^1d\mu \frac{\epsilon^2
p^2}{\tau}\frac{1}{P}\frac{\partial{g}}{\partial{p'}}\right).
\label{Schlickeiser2007-2-scaleanalysis}
\end{eqnarray}
With $\partial{}/\partial{z'}\sim O(1)$,
$\partial{}/\partial{t'}\sim O(1)$,
$\partial{}/\partial{p'}\sim O(1)$,
and
$\partial{}/\partial{\mu}\sim O(1)$,
the magnitude of 
the two terms on the left hand side of 
Equation (\ref{Schlickeiser2007-2-scaleanalysis})
are $O(1)$, and the magnitude of the 
first, second, and third terms
on the right hand side are $O(1)$, $O(1)$, and 
$\epsilon$, respectively. 

In the following subsections, we find 
that the momentum diffusion coefficient 
$A(L)$, i.e., $A(\xi)$,
can be written as the linear expression of 
$\xi^n\epsilon^m$ with 
$n,m=0,1,2,3,\cdots$
(see Appendix \ref{Evaluating A}).
The general form of $A(\xi)$ 
can be shown as 
\begin{eqnarray}
A(L)=A(\xi)=A+A'+\mathcal{M}(\epsilon, \xi)
\label{A(L)-general form-0} 
\end{eqnarray}
with
\begin{eqnarray}
\mathcal{M}(\epsilon,\xi)=M_1(\epsilon)\xi+M_2(\epsilon)\xi^2+M_3(\epsilon)\xi^3+\cdots
=\sum_{n=1}^{\infty}M_n(\epsilon)\xi^n.
\label{Mxi-0}
\end{eqnarray}
Here, $A$ and $A'$ are the second- and the higher-order of 
the momentum diffusion coefficient of 
the SOMT for constant field, 
and $\mathcal{M}(\epsilon, \xi)$ is the 
coefficient of the SOMT 
contributed from focusing effect. 
If $\mathcal{M}(\epsilon, \xi)\ne 0$, it indicates that the influence of
focusing effect to the momentum diffusion exists.  
The coefficients in Equation (\ref{Mxi-0})
are given as follows
\begin{eqnarray}
&&M_1(\epsilon)=M_{10}+M_{11}\epsilon+M_{12}\epsilon^2
+M_{13}\epsilon^3+\cdots
=\sum_{p=0}^{\infty}M_{1p}\epsilon^p,
\label{M1}\\
&&M_2(\epsilon)=M_{20}+M_{21}\epsilon+M_{22}\epsilon^2
+M_{23}\epsilon^3+\cdots
=\sum_{p=0}^{\infty}M_{2p}\epsilon^p,\\
&&\cdots \nonumber\\
&&M_n(\epsilon)=M_{n0}+M_{n1}\epsilon+M_{n2}\epsilon^2
+M_{n3}\epsilon^3+\cdots
=\sum_{p=0}^{\infty}M_{np}\epsilon^p,\\
\label{Mn}
&&\cdots \nonumber
\end{eqnarray}
with $p=1,2,3,\cdots$ and $q=0,1,2,\cdots$.  
Because $\xi\sim\epsilon$ is set in the above part, we can use 
$\eta$ to uniformly represent small parameters 
$\xi$ and $\epsilon$. 
Thus, Equation (\ref{A(L)-general form-0}) can be rewritten as
\begin{eqnarray}
A(\xi)\sim D_1\eta +D_2\eta^2+D_3\eta^3+\cdots
=\sum_{n=1}^{\infty}D_n\eta^n,
\label{A(L)-eta}
\end{eqnarray}
where, $D_1$, $D_2$, $\cdots$ are the coefficients. 

In the following, for the sake of simplicity
we ignore all of the terms higher than $\eta^4$ 
and retain the terms with lower or equal to 
$\eta^4$, i.e.,
\begin{eqnarray}
A_4(\xi)\sim D_1\eta +D_2\eta^2+D_3\eta^3+D_4\eta^4.
\label{A(L)-eta-4}
\end{eqnarray}
With Equations (\ref{M1})-(\ref{Mn}),
the latter equation contains the first-order terms
$M_{10}\xi$, 
the second-order ones $M_{11}\epsilon\xi$ and $M_{20}\xi^2$,
the third-order ones
$M_{30}\xi^3$, $M_{21}\epsilon\xi^2$, and
$M_{12}\epsilon^2\xi$, 
and the fourth-order ones 
$M_{40}\xi^4$,
$M_{31}\epsilon\xi^3$, $M_{22}\epsilon^2\xi^2$, and
$M_{13}\epsilon^3\xi$.
Thus, the coefficient of the SOMT for focused field
exact up to fourth-order 
can be written as
\begin{eqnarray}
\mathcal{M}_4(\epsilon,\xi)=&&M_1\xi+M_2\xi^2+M_3\xi^3+M_4\xi^4
\label{Mxi-4}
\end{eqnarray}
with 
\begin{eqnarray}
&&M_1=M_{10}+M_{11}\epsilon+M_{12}\epsilon^2
+M_{13}\epsilon^3,
\label{M1-0}\\
&&M_2=M_{20}+M_{21}\epsilon+M_{22}\epsilon^2,
\label{M2-0}\\
&&M_3=M_{30}+M_{31}\epsilon,
\label{M3-0}\\
&&M_4=M_{40}.
\label{M4-0}
\end{eqnarray}

In order to derive the formula of the momentum diffusion coefficient $A(\xi)$,
from Equation (\ref{continuity equation-constant D for focusing field})
we find that $\partial{g}/\partial{\mu}$ and $\partial{g}/\partial{p}$ 
have to be obtained. 
Because we retain $A(\xi)$ exact up to the fourth order, i.e., $\eta^4$,
considering Equation 
(\ref{continuity equation-constant D for focusing field}),
we retain $\partial{g}/\partial{\mu}$ and  
$\partial{g}/\partial{p}$ up to
the third- and second-order, respectively.
In this paper, we only explore the influence of focusing effect on
the momentum diffusion, 
so the terms without focusing effect 
are not investigated. 
Consequently, we only retain the terms
containing $\xi$ in $\partial{g}/\partial{\mu}$ and
$\partial{g}/\partial{p}$. 
Thus, the terms of $\partial{g}/\partial{\mu}$ and
$\partial{g}/\partial{p}$,
which we will retain,
should contain at least the first order in $\xi^1$.
Therefore, 
we only need to retain $\partial{g}/\partial{\mu}$ and
$\partial{g}/\partial{p}$ exact up to
the second- and first-orders in small quantity $\eta$. 
For this case, the momentum diffusion coefficient 
exact up to the fourth-order in small quantity $\eta$ can be written as
\begin{eqnarray}
A_4(\xi)=A+A'+\mathcal{M}_4(\epsilon, \xi).
\label{A4-simplified general form} 
\end{eqnarray}
It is noted that some terms not containing $\xi$ is not included in
the latter equation.   

\subsection{The formulas of the quantities used in 
derivation of $\partial{g}/\partial{\mu}$ and 
$\partial{g}/\partial{p}$} 
\label{The formulas of the related quantities used in 
derivation process of g/mu}

Employing the anisotropic distribution function 
$g(z, \mu, p, t)$ (see Equation (\ref{g-focusing field})),
we can obtain
\begin{eqnarray}
&&\frac{\partial{g}}{\partial{\mu}}
=\left(F-L\frac{\partial{F}}{\partial{z}}\right)
\frac{2\partial{e^{M(\mu)}}/\partial{\mu}}{\int_{-1}^{1}
d\mu e^{M(\mu) }}
+\frac{\partial{e^{M(\mu)}}}{\partial{\mu}}\left[R(z, \mu, p, t)
-\frac{\int_{-1}^{1}d\mu
e^{M(\mu)}R(z, \mu, p, t)}
{\int_{-1}^{1}d\mu
e^{M(\mu) }}\right]
+e^{M(\mu)}\frac{\partial{R}}{\partial{\mu}}.
\label{g/mu-focusing field-0}
\end{eqnarray}
Because the first term on the right-hand side of the latter 
equation does not contribute to $\partial{F}/\partial{p}$, 
using the simplification rule in 
Subsection (\ref{The lowest order momentum diffusion coefficient}), 
we can simplify Equation (\ref{g/mu-focusing field-0}) 
as
\begin{eqnarray}
\frac{\partial{g}}{\partial{\mu}}&&\Longrightarrow
\frac{\partial{e^{M(\mu)}}}{\partial{\mu}}\left[R(z, \mu, p, t)
-\frac{\int_{-1}^{1}d\mu
e^{M(\mu)}R(z, \mu, p, t)}
{\int_{-1}^{1}d\mu
e^{M(\mu) }}\right]
+e^{M(\mu)}\frac{\partial{R}}{\partial{\mu}},
\label{g/mu-focusing field-1}
\end{eqnarray}
where $R(z, \mu, p, t)$ and $\partial{R(z, \mu, p, t)}/\partial{\mu}$ need to be derived. 
Combinging 
Equations (\ref{Phi}) and (\ref{R}) gives
\begin{eqnarray}
R(z, \mu, p, t)=&&\int_{-1}^\mu d\nu \frac{e^{-M(\nu)}}{D_{\nu\nu}(\nu)} 
\Bigg[\frac{\partial{F}}{\partial{t}}\nu+\frac{\partial{}}{\partial{t}}
\int_{-1}^{\nu}d\rho g
+v\frac{\partial{}}{\partial{z}}\int_{-1}^{\nu}d\rho\rho g -\frac{v}{2} 
\frac{\partial{}}{\partial{z}}\int_{-1}^1d\mu\mu g\nonumber\\
&&-D_{\nu p}\frac{\partial{F}}{\partial{p}}
-D_{\nu p}\frac{\partial{g}}{\partial{p}}
-\frac{1}{p^2}\frac{\partial{}}{\partial{p}}p^2
\int_{-1}^\nu d\rho\left(D_{p\rho}\frac{\partial{g}}{\partial{\rho}}
+D_{pp}\frac{\partial{F}}{\partial{p}}
+D_{pp}\frac{\partial{g}}{\partial{p}}\right)
\nonumber\\
&&+\frac{1}{p^2}\frac{\partial{}}{\partial{p}}p^2
\frac{1}{2}\int_{-1}^1d\mu
\left(D_{pp}\frac{\partial{F}}{\partial{p}}
+D_{p\mu}\frac{\partial{g}}{\partial{\mu}}
+D_{pp}\frac{\partial{g}}{\partial{p}}
\right)\Bigg].
\end{eqnarray}
To ignore the terms not contributing to 
$\partial{F}/\partial{p}$ one obtains 
\begin{eqnarray}
R(z, \mu, p, t)&&\Longrightarrow-\frac{\partial{F}}{\partial{p}}
\int_{-1}^\mu d\nu e^{-M(\nu)}
\frac{D_{\nu p}}{D_{\nu\nu}}\textcircled{1}
-\frac{\partial{F}}{\partial{p}}
\int_{-1}^\mu d\nu \frac{e^{-M(\nu)}}{D_{\nu\nu}(\nu)}
\int_{-1}^\nu d\rho\frac{1}{p^2}\frac{\partial{}}{\partial{p}}\left(p^2D_{pp}\right)
\textcircled{2}\nonumber\\
&&+\frac{\partial{F}}{\partial{p}}
\int_{-1}^\mu d\nu \frac{e^{-M(\nu)}}{D_{\nu\nu}(\nu)}
\frac{1}{2}\int_{-1}^1d\mu\frac{1}{p^2}\frac{\partial{}}{\partial{p}}
\left(p^2D_{pp}\right)\textcircled{2}
-\int_{-1}^\mu d\nu e^{-M(\nu)}
\frac{D_{\nu p}}{D_{\nu\nu}(\nu)}\textcircled{1}
\frac{\partial{g}}{\partial{p}}
\nonumber\\
&&-
\int_{-1}^\mu d\nu \frac{e^{-M(\nu)}}{D_{\nu\nu}(\nu)}
\int_{-1}^\nu d\rho\frac{\partial{g}}{\partial{\rho}}\frac{1}{p^2}
\frac{\partial{}}{\partial{p}}
\left(p^2D_{p\rho}\right)\textcircled{1}
-\int_{-1}^\mu d\nu \frac{e^{-M(\nu)}}{D_{\nu\nu}(\nu)}
\int_{-1}^\nu d\rho D_{p\rho}\textcircled{1}
\frac{\partial{}}{\partial{p}}
\left(\frac{\partial{g}}{\partial{\rho}}
\right)\nonumber\\
&&-\int_{-1}^\mu d\nu \frac{e^{-M(\nu)}}{D_{\nu\nu}(\nu)}
\int_{-1}^\nu d\rho\frac{\partial{g}}{\partial{p}}\frac{1}{p^2}
\frac{\partial{}}{\partial{p}}
\left(p^2D_{pp}\right)\textcircled{2}
-\int_{-1}^\mu d\nu \frac{e^{-M(\nu)}}{D_{\nu\nu}(\nu)}
\int_{-1}^\nu d\rho
D_{pp}\frac{\partial{}}{\partial{p}}
\left(\frac{\partial{g}}{\partial{p}}\right)\textcircled{2}
\nonumber\\
&&+
\int_{-1}^\mu d\nu \frac{e^{-M(\nu)}}{D_{\nu\nu}(\nu)}
\frac{1}{2}\int_{-1}^1d\mu
\frac{\partial{g}}{\partial{\mu}}\frac{1}{p^2}
\frac{\partial{}}{\partial{p}}
\left(p^2D_{p\mu}\right)\textcircled{1}
+\int_{-1}^\mu d\nu \frac{e^{-M(\nu)}}{D_{\nu\nu}(\nu)}
\frac{1}{2}\int_{-1}^1d\mu D_{p\mu}
\frac{\partial{}}{\partial{p}}
\left(\frac{\partial{g}}{\partial{\mu}}
\right)\textcircled{1}\nonumber\\
&&+\int_{-1}^\mu d\nu \frac{e^{-M(\nu)}}{D_{\nu\nu}(\nu)}
\frac{1}{2}\int_{-1}^1d\mu
\frac{\partial{g}}{\partial{p}}\frac{1}{p^2}\frac{\partial{}}{\partial{p}}
\left(p^2D_{pp}\right)\textcircled{2}
+\int_{-1}^\mu d\nu \frac{e^{-M(\nu)}}{D_{\nu\nu}(\nu)}
\frac{1}{2}\int_{-1}^1d\mu 
D_{pp}\frac{\partial{}}{\partial{p}}
\left(\frac{\partial{g}}{\partial{p}}\right)\textcircled{2},
\label{R-1}
\end{eqnarray}
where the orders of small quantity 
$\epsilon$ are marked by the numbers circled. 
Furthermore, $\partial{R}/\partial{\mu}$
can be found as
\begin{eqnarray}
\frac{\partial{R}}{\partial{\mu}}
=&&-\frac{\partial{F}}{\partial{p}}
e^{-M(\mu)}
\frac{D_{\mu p}}{D_{\mu\mu}}\textcircled{1}
-\frac{\partial{F}}{\partial{p}}
\frac{e^{-M(\mu)}}{D_{\mu\mu}(\mu)}
\int_{-1}^\mu d\nu\frac{1}{p^2}\frac{\partial{}}{\partial{p}}\left(p^2D_{pp}\right)
\textcircled{2}
+\frac{\partial{F}}{\partial{p}}
\frac{e^{-M(\mu)}}{D_{\mu\mu}(\mu)}
\frac{1}{2}\int_{-1}^1d\mu\frac{1}{p^2}\frac{\partial{}}{\partial{p}}\left(p^2D_{pp}\right)\textcircled{2}
\nonumber\\
&&-e^{-M(\mu)}
\frac{D_{\mu p}}{D_{\mu\mu}(\mu)}\textcircled{1}
\frac{\partial{g}}{\partial{p}}
-
\frac{e^{-M(\mu)}}{D_{\mu\mu}(\mu)}
\int_{-1}^\mu d\nu\frac{\partial{g}}{\partial{\nu}}\frac{1}{p^2}
\frac{\partial{}}{\partial{p}}
\left(p^2D_{p\nu}\right)\textcircled{1}
-\frac{e^{-M(\mu)}}{D_{\mu\mu}(\mu)}
\int_{-1}^\mu d\nu D_{p\nu}\textcircled{1}
\frac{\partial{}}{\partial{p}}
\left(\frac{\partial{g}}{\partial{\nu}}
\right)
\nonumber\\
&&-\frac{e^{-M(\mu)}}{D_{\mu\mu}(\mu)}
\int_{-1}^\mu d\nu\frac{\partial{g}}{\partial{p}}\frac{1}{p^2}
\frac{\partial{}}{\partial{p}}
\left(p^2D_{pp}\right)\textcircled{2}
-\frac{e^{-M(\mu)}}{D_{\mu\mu}(\mu)}
\int_{-1}^\mu d\nu
D_{pp}\frac{\partial{}}{\partial{p}}
\left(\frac{\partial{g}}{\partial{p}}\right)\textcircled{2}
\nonumber\\
&&+\frac{e^{-M(\mu)}}{D_{\mu\mu}(\mu)}
\frac{1}{2}\int_{-1}^1d\mu
\frac{\partial{g}}{\partial{\mu}}\frac{1}{p^2}
\frac{\partial{}}{\partial{p}}
\left(p^2D_{p\mu}\right)\textcircled{1}
+\frac{e^{-M(\mu)}}{D_{\mu\mu}(\mu)}
\frac{1}{2}\int_{-1}^1d\mu D_{p\mu}
\frac{\partial{}}{\partial{p}}
\left(\frac{\partial{g}}{\partial{\mu}}
\right)\textcircled{1}
\nonumber\\
&&+\frac{e^{-M(\mu)}}{D_{\mu\mu}(\mu)}
\frac{1}{2}\int_{-1}^1d\mu
\frac{\partial{g}}{\partial{p}}\frac{1}{p^2}\frac{\partial{}}{\partial{p}}
\left(p^2D_{pp}\right)\textcircled{2}
+\frac{e^{-M(\mu)}}{D_{\mu\mu}(\mu)}
\frac{1}{2}\int_{-1}^1d\mu 
D_{pp}\frac{\partial{}}{\partial{p}}
\left(\frac{\partial{g}}{\partial{p}}\right)
\textcircled{2}.
\label{R/mu-1}
\end{eqnarray}

From Equation (\ref{R-1}), 
we find that $\partial{g}/\partial{p}$, 
$\partial^2{g}/(\partial{p}\partial{\mu})$, and 
$\partial^2{g}/\partial{p^2}$ have to be deduced.
Analogous to the derivation in the above, 
the formula of $\partial{g}/\partial{p}$ can be found as
\begin{eqnarray}
\frac{\partial{g}}{\partial{p}}
\Longrightarrow&&
\frac{\partial{F}}{\partial{p}}
\left(\frac{2e^{M(\mu)}}{\int_{-1}^{1}
d\mu e^{M(\mu) }}-1\right) 
+e^{M(\mu)}\frac{\partial{R}}{\partial{p}}
-e^{M(\mu)}\frac{1}{\int_{-1}^{1}d\mu
e^{M(\mu) }}\int_{-1}^{1}d\mu
e^{M(\mu)}\frac{\partial{R}}{\partial{p}}.
\label{g/p-1}
\end{eqnarray}
Here, $\partial{R}/\partial{p}$ can be also derived 
and are shown in Appendix \ref{The formula of
dR/dp}. The formula of $\partial{R}/\partial{p}$
shows that  $\partial^2{g}/(\partial{p}\partial{\mu})$,
$\partial^2{g}/\partial{p^2}$, $\partial^3{g}/(\partial{p^2}\partial{\mu})$, and 
$\partial^3{g}/\partial{p^3}$ have to be obtained.
We can easily find 
\begin{eqnarray}
&&\frac{\partial{}}{\partial{p}}\frac{\partial{g}}{\partial{\mu}}
\Longrightarrow
\frac{\partial{F}}{\partial{p}}
\frac{2\frac{\partial{e^{M(\mu)}}}{\partial{\mu}}}{\int_{-1}^{1}d\mu e^{M(\mu)}}
+\frac{\partial{e^{M(\mu)}}}{\partial{\mu}}\frac{\partial{}}{\partial{p}}R(\mu,t)
-\frac{\partial{e^{M(\mu)}}}{\partial{\mu}}\frac{\int_{-1}^{1}d\mu
e^{M(\mu)}\frac{\partial{}}{\partial{p}}R(\mu,t)}
{\int_{-1}^{1}d\mu e^{M(\mu) }}
+e^{M(\mu)}
\frac{\partial{}}{\partial{p}}
\frac{\partial{}}{\partial{\mu}}R(\mu,t),\\
&&\frac{\partial{}}{\partial{p}}\frac{\partial{g}}{\partial{p}}
\Longrightarrow
e^{M(\mu)}\frac{\partial{}}{\partial{p}}\frac{\partial{}}{\partial{p}}R(\mu,t)
-\frac{e^{M(\mu)}}{\int_{-1}^{1}d\mu e^{M(\mu) }}
\int_{-1}^{1}d\mu
e^{M(\mu)}
\frac{\partial{}}{\partial{p}}
\frac{\partial{}}{\partial{p}}R(\mu,t),\\
&&\frac{\partial{}}{\partial{p}}\frac{\partial{}}{\partial{p}}
\frac{\partial{g}}{\partial{\mu}}
\Longrightarrow
\frac{\partial{e^{M(\mu)}}}{\partial{\mu}}
\frac{\partial{}}{\partial{p}}\frac{\partial{}}{\partial{p}}R(\mu,t)
-\frac{\partial{e^{M(\mu)}}}{\partial{\mu}}\frac{\int_{-1}^{1}d\mu 
e^{M(\mu)}\frac{\partial{}}{\partial{p}}
\frac{\partial{}}{\partial{p}}R(\mu,t)}
{\int_{-1}^{1}d\mu e^{M(\mu) }}
+e^{M(\mu)}
\frac{\partial{}}{\partial{p}}\frac{\partial{}}{\partial{p}}
\frac{\partial{}}{\partial{\mu}}R(\mu,t),\\
&&\frac{\partial{}}{\partial{p}}
\frac{\partial{}}{\partial{p}}\frac{\partial{g}}{\partial{p}}
\Longrightarrow
e^{M(\mu)}
\frac{\partial{}}{\partial{p}}\frac{\partial{}}{\partial{p}}
\frac{\partial{}}{\partial{p}}R(\mu,t)
-\frac{e^{M(\mu)}}{\int_{-1}^{1}d\mu e^{M(\mu) }}
\int_{-1}^{1}d\mu
e^{M(\mu)}
\frac{\partial{}}{\partial{p}}\frac{\partial{}}{\partial{p}}
\frac{\partial{}}{\partial{p}}R(\mu,t).
\end{eqnarray}
Here appear new quantities
$\partial^2{R}/\partial{p^2}$,
$\partial^2{R}/(\partial{p}\partial{\mu})$,
and $\partial^3{R}/\partial{p^3}$.
In the following deduction, 
we can find that 
$\partial^3{R}/\partial{p^3}$ is not 
actually used.
So we just need to derive $\partial^2{R}/\partial{p^2}$ and
$\partial^2{R}/(\partial{p}\partial{\mu})$,
which is shown in Appendix 
\ref{The formulas ofd2R/dp2 and d2R/dpmu}. 
Here, the quantities not contributing to
the coefficient of the SOMT are not derived.  

\subsection{The formulas of $\partial{g}/\partial{\mu}$
and $\partial{g}/\partial{p}$}
\label{The formulas of g/mu and g/p}

By employing the iteration method \citep{wq2018, wq2019,
wq2020} and the simplification rule in 
Subsection (\ref{The lowest order momentum diffusion coefficient}), 
starting from Equation (\ref{g/mu-focusing field-1}) we can obtain the formulas of 
$\partial{g}/\partial{\mu}$ and 
$\partial{g}/\partial{p}$. 
Because the formulas of 
$\partial{g}/\partial{\mu}$ and 
$\partial{g}/\partial{p}$
are lenghty and complicated, we put them 
in the Appendix \ref{The formula of g/mu} 
and \ref{The formula of g/p}, respectively.

\subsection{The formula of the momentum diffusion coefficient $A(L)$}
\label{The formula of the momentum diffusion coefficient A  with the adiabatic focusing effect}

Inserting Equations (\ref{g/mu-10}) and (\ref{g/p-10})
into the right hand side of 
Equation (\ref{continuity equation-constant D for focusing field}), 
we can obtain the formula of $A_4(\xi)$ as
\begin{eqnarray}
A_4(\xi)=A+A'+\mathcal{M}_4(\xi)=\sum_{n=1}^{253}T_n,
\label{A4-Tn}
\end{eqnarray}
where all of the terms $T_n$ higher than fourth-order 
are set to zero, and the other ones
are evaluated and listed in Appendix 
\ref{Evaluating A}.
From the results of evaluation,
we can find that $A_4(\xi)$ is the linear function of 
$\xi^n\epsilon^m$ with 
$n,m=0,1,2,3,4$.

\section{Evaluating the momentum diffusion coefficient with the focusing effect}
\label{Evaluating the momentum diffusion coefficient $A$  with the adiabatic focusing effect}

To sum all of the terms $T_n$ in Equation (\ref{A4-Tn}),
from Equations (\ref{Mxi-4})-(\ref{M4-0})
we find
\begin{eqnarray}
&&M_{10}=M_{11}=M_{12}=0,\\
&&M_{13}=\frac{2}{9}p^2 D_2 \left(7 
-20\frac{D_2^2}{D_1^2}\right)\\
&&M_{20}=M_{21}=0,\\
&&M_{22}=p^2D_1  \left(\frac{188}{135}\frac{D_2^2}{D_1^2}
-\frac{2}{45} \right),\\
&&M_{30}=M_{31}=0,\\
&&M_{40}=0.
\end{eqnarray}
Therefore, we obtain
\begin{eqnarray}
&&M_1=\frac{2}{9}\epsilon^3 p^2 D_2 \left(7 
-20\frac{D_2^2}{D_1^2}\right),
\label{M1-result}\\
&&M_2=\epsilon^2 p^2D_1  \left(\frac{188}{135}\frac{D_2^2}{D_1^2}
-\frac{2}{45} \right),
\label{M2-result}\\
&&M_3=0,
\label{M3-result} \\
&&M_4=0.
\label{M4-result}
\end{eqnarray}
The Equations (\ref{M1-result})-(\ref{M4-result})
demonstrate that only the terms with $\epsilon^3\xi\sim\eta^4$ 
and $\epsilon^2\xi^2\sim\eta^4$ are not equal to zero, the other terms, 
that is, the ones with
$\xi\sim\eta$, $\epsilon\xi\sim\eta^2$, 
$\epsilon^2\xi\sim\eta^3$,
$\xi^2\sim\eta^2$, $\epsilon\xi^2\sim\eta^3$,
$\xi^3\sim\eta^3$, $\epsilon\xi^3\sim\eta^4$,
and $\xi^4\sim\eta^4$ 
are all equal to zero. 
Consequently, for the Fokker-Planck coefficients used in this paper (see Equations 
(\ref{Dmumu-2})-(\ref{Dpp-2})), 
the lowest order of $\mathcal{M}_4(\xi)$
is $\eta^4$.

Inserting Equations (\ref{M1-result})-(\ref{M4-result})
into Equation (\ref{Mxi-4}) gives
\begin{eqnarray}
\mathcal{M}_4(\epsilon,\xi)&&=D_1\left[\frac{2}{9}\epsilon^3  
\frac{D_2}{D_1} \left(7 
-20\frac{D_2^2}{D_1^2}\right)\xi+\epsilon^2  \left(\frac{188}{135}\frac{D_2^2}{D_1^2}
-\frac{2}{45} \right)\xi^2\right]p^2
\nonumber\\
&&
=D_1\left[\frac{2}{9}\epsilon^3  
H_C \left(7 
-20H_C^2\right)\xi+\epsilon^2  \left(\frac{188}{135}H_C^2
-\frac{2}{45} \right)\xi^2\right]p^2,
\label{Mxi-4-specific-0}
\end{eqnarray}
which indicates that the sign of 
$\mathcal{M}_4(\epsilon,\xi)$ is determined by 
$H_C$, and $\xi$. 
For the divergent field, i.e., $\xi>0$, using 
$\xi\sim\epsilon\sim\eta$ we obtain
\begin{eqnarray}
\mathcal{M}_4(\epsilon,\xi)
\sim D_1\left(-\frac{2}{45}+\frac{14}{9}H_C
+\frac{188}{135}H_C^2
-\frac{40}{9}H_C^3
\right)p^2\eta^4.
\label{Mxi-4-specific}
\end{eqnarray}
Because of $D_1>0$, 
the coefficient $\mathcal{M}_4(\epsilon,\xi)$
is negative for $H_C=0$ and $H_C= 1$. However, 
if $H_C= -1$, 
$\mathcal{M}_4(\epsilon,\xi)$ is positive.
For the convergent field, i.e., $\xi<0$, 
Equation (\ref{Mxi-4-specific-0}) becomes
\begin{eqnarray}
\mathcal{M}_4(\epsilon,\xi)\sim
D_1\left[-\frac{2}{45}-\frac{14}{9}H_C
+\frac{188}{135}H_C^2
+\frac{40}{9} H_C^3
\right]p^2\eta^4.
\label{Mxi-4-specific-for xi<0}
\end{eqnarray} 
For $H_C=0$ and $H_C=-1$
we can find that $\mathcal{M}_4(\epsilon,\xi)$ is negative 
and for $H_C=1$ the coefficient $\mathcal{M}_4(\epsilon,\xi)$ is
positive. 
From the above discussion we find that 
the sign of $\mathcal{M}_4(\epsilon,\xi)$
is determined by the parameters $H_C$ and $\xi$. 
The above results are listed in table \ref{The modifying term sign}. 

\section{Discussion}
\label{Discussion}

In Subection \ref{The isotropic distribution function equation for constant field}, we obtain the EIDF for the 
constant field. For the convenience of comparison,
here we write it down again   
\begin{eqnarray}
\frac{\partial{F}}{\partial{t}}
&&=\frac{1}{p^2}\frac{\partial{}}{\partial{p}}
\left(p^2
A
\frac{\partial{F}}{\partial{p}}\right)
+\frac{1}{p^2}\frac{\partial{}}{\partial{p}}
\left(p^2
A'
\frac{\partial{F}}{\partial{p}}\right)
+\frac{1}{p^2}\frac{\partial{}}{\partial{p}}
\Bigg[p^2\kappa_{3p1}^c
\frac{\partial{}}{\partial{p}}
\left(\kappa_{3p2}^c
\frac{\partial{F}}{\partial{p}}\right)\Bigg]
+\frac{1}{p^2}\frac{\partial{}}{\partial{p}}
\Bigg\{p^2
\kappa_{4p1}^c\frac{\partial{}}{\partial{p}}
\Bigg[\kappa_{4p2}^c\frac{\partial{}}{\partial{p}}
\left(\kappa_{4p3}^c\frac{\partial{F}}{\partial{p}}
\right)\Bigg]\Bigg\}
+\cdots,
\label{equation of F for constant field-2}
\end{eqnarray}
which contains the coefficients $A$, $A'$,
$\kappa_{3p1}^c$, $\kappa_{3p2}^c$, $\cdots$ 
for the uniform field.
From the investigation in the previous sections,
the EIDF for the focused field can be obtained as
\begin{eqnarray}
\frac{\partial{F}}{\partial{t}}
=&&
\frac{1}{p^2}\frac{\partial{}}{\partial{p}}
\Bigg(p^2\kappa_p^f F\Bigg)
+\frac{1}{p^2}\frac{\partial{}}{\partial{p}}
\Bigg(p^2
A\frac{\partial{F}}{\partial{p}}
\Bigg)
+\frac{1}{p^2}\frac{\partial{}}{\partial{p}}
\left(p^2
A'
\frac{\partial{F}}{\partial{p}}\right)
+\frac{1}{p^2}\frac{\partial{}}{\partial{p}}
\Bigg(p^2
\mathcal{M}_4(\epsilon,\xi)\frac{\partial{F}}{\partial{p}}
\Bigg)
\nonumber\\
&&
+\frac{1}{p^2}\frac{\partial{}}{\partial{p}}
\Bigg[p^2\kappa_{3p1}^f
\frac{\partial{}}{\partial{p}}
\left(\kappa_{3p2}^f
\frac{\partial{F}}{\partial{p}}\right)\Bigg]
+\cdots
\label{equation of F for focusing field}
\end{eqnarray}
with the coefficient of the SOMT for 
constant field, $A$ and $A'$, and the coefficient of 
the SOMT for the focused field, 
$\mathcal{M}_4(\epsilon,\xi)$.
Comparing Equations (\ref{equation of F for constant field-2}) with
(\ref{equation of F for focusing field}), 
we can find that the focused field contributes to
a momentum streaming term, i.e., the first term
on the right hand side of Equation
(\ref{equation of F for focusing field}),
which was explored
by \citet{SchlickeiserEA2008} and 
\citet{LitvinenkoEA2011}.
Moreover, an additional 
SOMT occurs 
in Equation (\ref{equation of F for focusing field}),
obviously, which 
is also caused by the focused field.  
In fact, using the same operation in this paper, 
we can find that the adiabatic focusing effect can 
affect the higher-order momentum derivative terms.
For example, the coefficients of the 
third-order momentum derivative term of Equation 
(\ref{equation of F for focusing field}), $\kappa_{3p1}^f$
and $\kappa_{3p2}^f$, are not equal to 
the coefficients $\kappa_{3p1}^c$
and $\kappa_{3p2}^c$ of 
Equation (\ref{equation of F for constant field-2}),
respectively.  
In this paper, we only explore the SOMT and will 
investigate the higher-order terms in the future. 

By employing Equations (\ref{equation of F for focusing field}) we can obtain the mean change rate of the particle momentum with time as
\begin{eqnarray}
\frac{\dee\langle p \rangle}{\dee t}&=&
\frac{\dee}{\dee t}\int dp p^3 F(p, t)
=\int dp p^3 
\frac{\partial{F}}{\partial{t}}
\nonumber\\
&=&-\left\langle \kappa_p \right\rangle
+\left\langle \frac{1}{p^2}\frac{\partial{}}{\partial{p}}
\left(p^2A_0 \right)\right\rangle
+\left\langle \frac{1}{p^2}\frac{\partial{}}{\partial{p}}
\left(p^2\mathcal{M}_4(\epsilon,\xi) \right)\right\rangle
-\left\langle \frac{1}{p^2}\frac{\partial{}}{\partial{p}}
\left(p^2\kappa_{3p2}^f \frac{\partial{\kappa_{3p1}^f}}{\partial{p}}\right)
\right\rangle
+\cdots.
\label{d<p>/dt}
\end{eqnarray}
One can find that the mean particle momentum is determined by 
all the coefficients of Equation 
(\ref{equation of F for focusing field}).
Equation (\ref{d<p>/dt}) can also be rewritten as 
\begin{eqnarray}
\frac{\dee\langle p \rangle}{\dee t}
&=&\left(\frac{\dee\langle p \rangle}{\dee t}\right)_1
+\left(\frac{\dee\langle p \rangle}{\dee t}\right)_{21}
+\left(\frac{\dee\langle p \rangle}{\dee t}\right)_{new}
+\left(\frac{\dee\langle p \rangle}{\dee t}\right)_3
+\cdots
\end{eqnarray}
with
\begin{eqnarray}
&&\left(\frac{\dee\langle p \rangle}{\dee t}\right)_1
=-\left\langle \kappa_p \right\rangle,\\
&&\left(\frac{\dee\langle p \rangle}{\dee t}\right)_{21}
=\left\langle \frac{1}{p^2}\frac{\partial{}}{\partial{p}}
\left(p^2A_0 \right)\right\rangle,\\
&&\left(\frac{\dee\langle p \rangle}{\dee t}\right)_{22}
=\left\langle \frac{1}{p^2}\frac{\partial{}}{\partial{p}}
\left(p^2\mathcal{M}_4(\epsilon,\xi) \right)\right\rangle,\\
&&\left(\frac{\dee\langle p \rangle}{\dee t}\right)_3
=-\left\langle \frac{1}{p^2}\frac{\partial{}}{\partial{p}}
\left(p^2\kappa_{3p2}^f \frac{\partial{\kappa_{3p1}^f}}{\partial{p}}\right)
\right\rangle,\\
&&\cdots.
\end{eqnarray}
In the following, we evaluate 
the mean momentum change rate 
$\left(\dee\langle p \rangle/\dee t\right)_{22}$ 
contributed from the additional SOMT, which is shown as 
\begin{eqnarray}
\left(\frac{\dee\langle p \rangle}{\dee t}\right)_{22}
=\left\langle \frac{1}{p^2}\frac{\partial{}}{\partial{p}}
\left(p^2\mathcal{M}_4(\epsilon,\xi) \right)\right\rangle
=4D_1\left[\frac{2}{9}\epsilon^3  
H_C \left(7 
-20H_C^2\right)\xi+\epsilon^2  \left(\frac{188}{135}H_C^2
-\frac{2}{45} \right)\xi^2\right]
\left\langle p\right\rangle,
\end{eqnarray}
where Equation (\ref{Mxi-4-specific-0}) is used. 
The solution of the latter equation can be found
\begin{eqnarray}
\langle p \rangle_{22}\sim \langle p \rangle_0
e^{4D_1\alpha \eta^4 t}
\label{<p>new}
\end{eqnarray}
with
\begin{eqnarray}
\alpha=\frac{2}{9}\epsilon^3  
H_C \left(7 
-20H_C^2\right)\xi+\epsilon^2  \left(\frac{188}{135}H_C^2
-\frac{2}{45} \right)\xi^2.
\label{alpha}
\end{eqnarray}
According to the analysis in Section 
\ref{Evaluating the momentum diffusion coefficient $A$  with the adiabatic focusing effect}, 
for most cases the quantity $\mathcal{M}_4(\epsilon,\xi)$
is not equal to zero, and the parameter 
$\alpha$ likewise. 
Obviously, so long as the third-order algebraic polynomial 
(\ref{alpha})
over the parameter $H_c$
is not equal to zero, the 
mean particle momentum
$\langle p \rangle_{22}$ varies with time. 
Therefore, the focused field provides 
an additional momentum loss or 
gain process. 
This physical process exists
as long as the background magnetic field is nonuniform.

By setting $p/P^*=P'$ with the dimensionless momentum $P'$
and the momentum characteristic quantity $P^*$,
and regulating $\partial{F}/\partial{P'}\sim \Delta F/\Delta P'\sim O(1)$
with $\Delta P'\sim O(1)$ and $\Delta F\sim O(1)$,  
we can rewrite the SOMT and momentum convective terms as
\begin{eqnarray}
\frac{1}{p^2}\frac{\partial{}}{\partial{p}}\left(p^2A
\frac{\partial{F}}{\partial{p}}\right)\sim
\frac{2}{3} \frac{p^2}{P^{*2}}D_1\left(1-H_C^2\right)\epsilon^2
\end{eqnarray}
and 
\begin{eqnarray}
\kappa_p\frac{\partial{F}}{\partial{p}}\sim
\frac{2}{3} \frac{p}{P^*} D_1H_C\epsilon\xi
\end{eqnarray}
with
\begin{eqnarray}
\kappa_p=\frac{v}{4L}\int_{-1}^1d\mu 
\frac{D_{p\mu}\left(1-\mu^2\right)}{D_{\mu\mu}}
=\frac{1}{3L} p H_Cv_A.
\label{kp}
\end{eqnarray}

If setting  $p/P^*=P'\sim O(1)$, we can find that 
\begin{eqnarray}
&&\frac{1}{p^2}\frac{\partial{}}{\partial{p}}\left(p^2A
\frac{\partial{F}}{\partial{p}}\right)\sim
\frac{2}{3} D_1\left(1-H_C^2\right)\epsilon^2
\end{eqnarray}
and 
\begin{eqnarray}
\kappa_p\frac{\partial{F}}{\partial{p}}\sim
\frac{2}{3} D_1H_C\epsilon\xi.
\end{eqnarray}

Because of $\xi\sim\epsilon$, the relation 
$A\sim\kappa_p$ can be obtained,
which means that $A$ and $\kappa_p$ are on the same order of 
the relative importance. 
If $\xi\gg\epsilon$, the momentum convective term is far larger than 
the SOMT, the case of which was explored by \citet{LitvinenkoEA2011}
and \citet{ArmstrongEA2012}. 
In contrast, for $\xi\ll\epsilon$
the momentum convective term
is much less than the SOMT.
Therefore, 
the SOMT should be considered 
for $\xi\sim\epsilon$ and $\xi\ll\epsilon$. 
Thus, for very weak focusing limit 
the SOMT cannot be ignored
if the momentum convective term
need to be considered. 

For energetic particles
experiencing a large number of converging and diverging background
magnetic fields, 
the change rate of the particle momentum 
caused by the first-order focusing acceleration effect  
is
\begin{eqnarray}
\left(\frac{\dee\langle p \rangle}{\dee t}\right)_{First}
=-\langle \kappa_p \rangle
=-\frac{1}{3L} H_Cv_A\langle p \rangle.
\end{eqnarray}
For almost the same 
number of converging and diverging background
magnetic fields,
the mean value of $(\dee\langle p \rangle/\dee t)_{First}$ 
is equal approximately to zero.
Similarily, 
the change rate of the particle momentum 
caused by the second-order focusing acceleration effect  
is
\begin{eqnarray}
\left(\frac{\dee\langle p \rangle}{\dee t}\right)_{22}
=\left\langle \frac{1}{p^2}\frac{\partial{}}{\partial{p}}
\left(p^2\mathcal{M}_4(\xi) \right)\right\rangle
\end{eqnarray}
with
\begin{eqnarray}
&&\mathcal{M}_4(\xi)=M_1\xi+M_2\xi^2,\\
&&M_1=\frac{2}{9}\epsilon^3 p^2 D_2 \left(7 
-20\frac{D_2^2}{D_1^2}\right),\\
&&M_2=\epsilon^2 p^2D_1  \left(\frac{188}{135}\frac{D_2^2}{D_1^2}
-\frac{2}{45} \right),
\end{eqnarray}
Here, $\xi=\lambda/L$. Obviously, 
the mean value of $(\dee\langle p \rangle/\dee t)_{22}$ for 
almost the same 
number of converging and diverging background
magnetic fields
becomes
\begin{eqnarray}
\left(\frac{\dee\langle p \rangle}{\dee t}\right)_{22}
=\left\langle \frac{1}{p^2}\frac{\partial{}}{\partial{p}}
\left(p^2M_2\right)\right\rangle\frac{\lambda^2}{L^2}
\ne 0,
\label{averaged second ordermomentum change rate caused by focused field}
\end{eqnarray}
which is not equal to zero for almost cases. 

\section{SUMMARY AND CONCLUSION}
\label{SUMMARY AND CONCLUSION}

In this paper, we explore
the momentum diffusion due to the 
along-field adiabatic focusing effect. 
By employing the iteration method
\citep{WangEA2017, wq2018, wq2019, wq2020}, 
from the Fokker-Planck equation with adiabatic focusing
effect 
we derive the equation of the isotropic distribution function (EIDF). Comparing with the EIDF for constant field,
we find that 
the EIDF for the focused field contains one additional  
first-order and one additional second-order 
momentum derivative terms, obviously, which are all
caused by the focused field. 
The first-order momentum derivative term was 
explored by \citet{SchlickeiserEA2008} and \citet{
LitvinenkoEA2011}. But the additional second-order 
momentum derivative term (SOMT) is new. Thereafter, 
we obtain the mean change rate of particle momentum 
$(d\langle p \rangle/dt)_{22}$ contributed from 
the additional SOMT.
We find that the quantity 
$(d\langle p \rangle/dt)_{22}$ is not equal to zero
for most case.
Thus, it is identified that 
the focused field provides an additional momentum loss or 
gain process through this additional SOMT. 
This physical process should exist 
as long as
the background magnetic field is nonuniform.
If the charged energetic particles go through almost 
the same number of
converging and diverging background
magnetic fields, the relation
$(d\langle p \rangle/dt)_{First}\approx 0$ 
can be found. 
However, for the same case, 
Equations (\ref{averaged second ordermomentum change rate caused by focused field}) indicates
the momentum change rate contributed from 
the additional SOMT is not equal to zero.

According to the scale analysis theory, 
as shown in the paper of \citet{GombosiEA1993}, 
all of the terms with the same order of relative importance should
be retained. 
In the present article, we assume
$\xi=\lambda/L$ and $\epsilon=V_A/v$ 
with the same order, i.e., 
$\xi\sim \epsilon$. For convenience  we use $\eta$ to 
represent the two small parameters $\xi$ and $\epsilon$.  
In order to explore detailly the properties of the 
SOMT,  
we retain the coefficient $\mathcal{M}(\epsilon,\xi)$
exact up to fourth order in $\eta$, and write it 
as $\mathcal{M}_4(\epsilon,\xi)$. 
For the divergent field, $\mathcal{M}_4(\epsilon,\xi)$
is simplified as Equation (\ref{Mxi-4-specific}). 
If $H_c=0$ and $H_c>0$, the coefficient $\mathcal{M}_4(\epsilon,\xi)$
is negative. Inserting into the formula of 
$(d\langle p \rangle/dt)_{22}$,
we can find that the mean momentum 
$\langle p \rangle_{22}$ deceases 
relative to that in uniform magnetic field.
However, if $H_c=-1$, the coefficient
$\mathcal{M}_4(\epsilon,\xi)$ is positive, 
the value of $\langle p \rangle_{new}$
shows the mean momentum increases.  
In addition, for the convergent field, 
if $H_c=0$ and $H_c=-1$ the coefficient $\mathcal{M}_4(\epsilon,\xi)$
is negative and the corresponding mean momentum 
$\langle p \rangle_{22}$ decreases. 
For $H_c=1$, the coefficient $\mathcal{M}_4(\epsilon,\xi)$ is positive and 
the corresponding mean momentum 
$\langle p \rangle_{22}$ increases. 
The above results show that the cross helicity $H_C$ 
determinds the sign of the 
coefficient $\mathcal{M}_4(\epsilon,\xi)$. 

According to the scale analysis 
theory we have to set the relation of the two small 
parameters $\xi$ and $\eta$.
For convenience, we set $\xi\sim\eta$ in this paper. 
If the two small 
parameters have other relative order relationship,
we will get the other results and
the derivation may becomes more complicated.
However, the main findings 
in this paper are still valid.  
Furthermore, we use 
the Fokker-Planck coefficients in 
\citet{Schlickeiser2002} which are
derived by the quasilinear theory.
This theory has been the 
standard approach for describing plasma and particle interactions in the last decades. 
However, it is discovered that 
quasilinear theory is problematic for many scenarios
in different research methods,  e.g.,
computer simulations. Therefore,
in order to more deeply 
explore the particle acceleration and transport,  
the improved theories or the more reasonable 
nonlinear theories for transport process have to 
be developed
in the future. 

The non-local transport, memory effect 
and non-Gaussian pdfs in the transport theory of 
energetic particles have been exploited 
in literature
\citep{NegreteEA2004, NegreteEA2005, BianEA2017}.  
Starting from the steady-state Fokker-Planck equation,
\citet{BianEA2017} derived the 
non-local flux-gradient expression which involves a convolution of the gradient of the isotropic distribution 
function with a Laplacian distribution. 
It is suggested that 
the particle transport equation 
could become an integro-differential equation,
which will be investigated in the future. 

\acknowledgments

The authors thank the anonymous referee for valuable comments.
We are partly supported by
grant  NNSFC 41874206 and NNSFC 42074206. 

\renewcommand{\theequation}{\Alph{section}-\arabic{equation}}
\begin{appendices}

\section{The formula of $\partial{R}/\partial{p}$}
\label{The formula of dR/dp}
\begin{eqnarray}
\frac{\partial{R}}{\partial{p}}
\Longrightarrow&&
-\frac{\partial{F}}{\partial{p}}\int_{-1}^\mu d\nu e^{-M(\nu)} 
\frac{\partial{}}{\partial{p}}\frac{D_{\nu p}}{D_{\nu\nu}(\nu)} 
\textcircled{1}
-\int_{-1}^\mu d\nu e^{-M(\nu)} \frac{D_{\nu p}}{D_{\nu\nu}(\nu)}
\textcircled{1} \frac{\partial{}}{\partial{p}}\frac{\partial{g}}{\partial{p}}
-\int_{-1}^\mu d\nu e^{-M(\nu)} 
\frac{\partial{g}}{\partial{p}}
\frac{\partial{}}{\partial{p}}
\left(\frac{D_{\nu p}}{D_{\nu\nu}(\nu)}\right)\textcircled{1} 
\nonumber\\
&&
-\int_{-1}^\mu d\nu \frac{e^{-M(\nu)}}{D_{\nu\nu}(\nu)}
\int_{-1}^\nu d\rho
\frac{\partial{g}}{\partial{\rho}}
\frac{\partial{}}{\partial{p}}
\left( \frac{1}{p^2}\frac{\partial{}}{\partial{p}}
\left(p^2D_{p\rho  }\right)\right)\textcircled{1}
-\int_{-1}^\mu d\nu \frac{e^{-M(\nu)}}{D_{\nu\nu}(\nu)}
\int_{-1}^\nu d\rho
\left(\frac{\partial{}}{\partial{p}}
\frac{\partial{g}}{\partial{\rho}}
\right)
\left( \frac{1}{p^2}\frac{\partial{}}{\partial{p}}
\left(p^2D_{p\rho}\right)\right)\textcircled{1}
\nonumber\\
&&
-\int_{-1}^\mu d\nu \frac{e^{-M(\nu)}}{D_{\nu\nu}(\nu)}
\int_{-1}^\nu d\rho
D_{p\rho}\textcircled{1}
\frac{\partial{}}{\partial{p}}
\frac{\partial{}}{\partial{p}}\frac{\partial{g}}{\partial{\rho}}
-\int_{-1}^\mu d\nu \frac{e^{-M(\nu)}}{D_{\nu\nu}(\nu)}
\int_{-1}^\nu d\rho
\frac{\partial{D_{p\rho}}}{\partial{p}}\textcircled{1}
\frac{\partial{}}{\partial{p}}\frac{\partial{g}}{\partial{\rho}}
\nonumber\\
&&
-\frac{\partial{F}}{\partial{p}}
\int_{-1}^\mu d\nu \frac{e^{-M(\nu)}}{D_{\nu\nu}(\nu)}
\int_{-1}^\nu d\rho
\frac{\partial{}}{\partial{p}}
\left( \frac{1}{p^2}\frac{\partial{}}{\partial{p}}
\left(p^2D_{pp}\right)\right)\textcircled{2}
-\int_{-1}^\mu d\nu \frac{e^{-M(\nu)}}{D_{\nu\nu}(\nu)}
\int_{-1}^\nu d\rho 
\frac{\partial{g}}{\partial{p}}
\frac{\partial{}}{\partial{p}}
\left(\frac{1}{p^2}\frac{\partial{}}{\partial{p}}
\left(p^2D_{pp}\right)\right)\textcircled{2}
\nonumber\\
&&
-\int_{-1}^\mu d\nu \frac{e^{-M(\nu)}}{D_{\nu\nu}(\nu)}
\int_{-1}^\nu d\rho
\left(\frac{\partial{}}{\partial{p}}\frac{\partial{g}}{\partial{p}}
\right)
\frac{1}{p^2}\frac{\partial{}}{\partial{p}}
\left(p^2D_{pp}\right)\textcircled{2}
-\int_{-1}^\mu d\nu \frac{e^{-M(\nu)}}{D_{\nu\nu}(\nu)}
\int_{-1}^\nu d\rho 
D_{pp}\textcircled{2}
\frac{\partial{}}{\partial{p}}
\frac{\partial{}}{\partial{p}}\frac{\partial{g}}{\partial{p}}
\nonumber\\
&&
-\int_{-1}^\mu d\nu \frac{e^{-M(\nu)}}{D_{\nu\nu}(\nu)}
\int_{-1}^\nu d\rho 
\frac{\partial{D_{pp}}}{\partial{p}}\textcircled{2}
\frac{\partial{}}{\partial{p}}\frac{\partial{g}}{\partial{p}}
+\frac{1}{2}\frac{\partial{F}}{\partial{p}}
\int_{-1}^\mu d\nu \frac{e^{-M(\nu)}}{D_{\nu\nu}(\nu)} 
\int_{-1}^1d\mu 
\frac{\partial{}}{\partial{p}}
\left(\frac{1}{p^2}\frac{\partial{}}{\partial{p}}
\left(p^2D_{pp}\right)\right)\textcircled{2}
\nonumber\\
&&
+\frac{1}{2}\int_{-1}^\mu d\nu \frac{e^{-M(\nu)}}{D_{\nu\nu}(\nu)}
\int_{-1}^1d\mu \frac{\partial{g}}{\partial{\mu}}
\frac{\partial{}}{\partial{p}}
\left(\frac{1}{p^2}\frac{\partial{}}{\partial{p}}
\left(p^2D_{p\mu}\right)\right)\textcircled{1}
+\frac{1}{2}\int_{-1}^\mu d\nu \frac{e^{-M(\nu)}}{D_{\nu\nu}(\nu)}
\int_{-1}^1d\mu 
\left(\frac{\partial{}}{\partial{p}}\frac{\partial{g}}{\partial{\mu}}
\right)
\left(\frac{1}{p^2}\frac{\partial{}}{\partial{p}}
\left(p^2D_{p\mu}\right)\right)\textcircled{1}
\nonumber\\
&&
+\frac{1}{2}\int_{-1}^\mu d\nu \frac{e^{-M(\nu)}}{D_{\nu\nu}(\nu)}
\int_{-1}^1d\mu 
D_{p\mu}\textcircled{1}
\frac{\partial{}}{\partial{p}}
\frac{\partial{}}{\partial{p}}\frac{\partial{g}}{\partial{\mu}}
+\frac{1}{2}\int_{-1}^\mu d\nu \frac{e^{-M(\nu)}}{D_{\nu\nu}(\nu)}
\int_{-1}^1d\mu 
\frac{\partial{D_{p\mu}}}{\partial{p}}\textcircled{1}
\frac{\partial{}}{\partial{p}}\frac{\partial{g}}{\partial{\mu}}
\nonumber\\
&&
+\frac{1}{2}\int_{-1}^\mu d\nu \frac{e^{-M(\nu)}}{D_{\nu\nu}(\nu)} 
\int_{-1}^1d\mu 
\frac{\partial{g}}{\partial{p}}
\frac{\partial{}}{\partial{p}}
\left(\frac{1}{p^2}\frac{\partial{}}{\partial{p}}
\left(p^2D_{pp}\right)\right)\textcircled{2}
+\frac{1}{2}\int_{-1}^\mu d\nu \frac{e^{-M(\nu)} }{D_{\nu\nu}(\nu)} 
\int_{-1}^1d\mu 
\left(\frac{\partial{}}{\partial{p}}\frac{\partial{g}}{\partial{p}}
\right)
\left(\frac{1}{p^2}\frac{\partial{}}{\partial{p}}
\left(p^2D_{pp}\right)\right)\textcircled{2}
\nonumber\\
&&
+\frac{1}{2}\int_{-1}^\mu d\nu  \frac{e^{-M(\nu)}}{D_{\nu\nu}(\nu)} 
\int_{-1}^1d\mu 
D_{pp}\textcircled{2}
\frac{\partial{}}{\partial{p}}
\frac{\partial{}}{\partial{p}}\frac{\partial{g}}{\partial{p}}
+\frac{1}{2}\int_{-1}^\mu d\nu  \frac{e^{-M(\nu)}}{D_{\nu\nu}(\nu)} 
\int_{-1}^1d\mu 
\frac{\partial{D_{pp}}}{\partial{p}}\textcircled{2}
\frac{\partial{}}{\partial{p}}\frac{\partial{g}}{\partial{p}}.
\label{R/p-1}
\end{eqnarray}

\section{The formulas of $\partial^2{R}/\partial{p^2}$ and
$\partial^2{R}/(\partial{p}\partial{\mu})$}
\label{The formulas ofd2R/dp2 and d2R/dpmu}
\begin{eqnarray}
&&\frac{\partial^2{R}}{\partial{p^2}}
\Longrightarrow 
-\frac{\partial{F}}{\partial{p}}
\int_{-1}^\mu d\nu e^{-M(\nu)} 
\frac{\partial{}}{\partial{p}}\frac{\partial{}}{\partial{p}}
\left(\frac{D_{\nu p}}{D_{\nu\nu}}\right) \textcircled{1} 
-
\int_{-1}^\mu d\nu e^{-M(\nu)} 
\frac{\partial{g}}{\partial{p}}
\frac{\partial{}}{\partial{p}}
\frac{\partial{}}{\partial{p}}\left(\frac{D_{\nu p}}{D_{\nu\nu}}\right)
\textcircled{1} 
\nonumber\\
&&-2\int_{-1}^\mu d\nu e^{-M(\nu)} 
\left(\frac{\partial{}}{\partial{p}}\frac{\partial{g}}{\partial{p}}
\right)
\frac{\partial{}}{\partial{p}}\left(\frac{D_{\nu p}}{D_{\nu\nu}}\right)
\textcircled{1} 
-\int_{-1}^\mu d\nu e^{-M(\nu)} 
\left(\frac{D_{\nu p}}{D_{\nu\nu}}\right)\textcircled{1}
\frac{\partial{}}{\partial{p}}
\frac{\partial{}}{\partial{p}}\frac{\partial{g}}{\partial{p}}
\nonumber\\
&&
-
\int_{-1}^\mu d\nu e^{-M(\nu)} \frac{1}{D_{\nu\nu}(\nu)}
\int_{-1}^\nu d\rho \frac{\partial{g}}{\partial{\rho}}
\frac{\partial{}}{\partial{p}}
\frac{\partial{}}{\partial{p}}\left(\frac{1}{p^2}\frac{\partial{}}{\partial{p}}
\left(p^2
D_{p\rho}\right)\right)\textcircled{1} 
\nonumber\\
&&
-
2\int_{-1}^\mu d\nu e^{-M(\nu)} \frac{1}{D_{\nu\nu}(\nu)}
\int_{-1}^\nu d\rho
\left(\frac{\partial{}}{\partial{p}}
\frac{\partial{g}}{\partial{\rho}}
\right)
\frac{\partial{}}{\partial{p}}\left(\frac{1}{p^2}\frac{\partial{}}{\partial{p}}
\left(p^2
D_{p\rho}\right)\right)\textcircled{1}
\nonumber\\
&&
-
\int_{-1}^\mu d\nu e^{-M(\nu)} \frac{1}{D_{\nu\nu}(\nu)}
\int_{-1}^\nu d\rho 
\left(\frac{\partial{}}{\partial{p}}\frac{\partial{}}{\partial{p}}
\frac{\partial{g}}{\partial{\rho}}
\right)
\left(\frac{1}{p^2}\frac{\partial{}}{\partial{p}}
\left(p^2
D_{p\rho}\right)\right)\textcircled{1} 
\nonumber\\
&&
-
\int_{-1}^\mu d\nu e^{-M(\nu)} \frac{1}{D_{\nu\nu}(\nu)}
\int_{-1}^\nu d\rho 
D_{p\rho}\textcircled{1}
\frac{\partial{}}{\partial{p}}
\frac{\partial{}}{\partial{p}}\frac{\partial{}}{\partial{p}}
\frac{\partial{g}}{\partial{\rho}} 
-
\int_{-1}^\mu d\nu e^{-M(\nu)} \frac{1}{D_{\nu\nu}(\nu)}
\int_{-1}^\nu d\rho 
\frac{\partial{D_{p\rho}}}{\partial{p}}\textcircled{1}
\frac{\partial{}}{\partial{p}}\frac{\partial{}}{\partial{p}}
\frac{\partial{g}}{\partial{\rho}} 
\nonumber\\
&&
-
\int_{-1}^\mu d\nu e^{-M(\nu)} \frac{1}{D_{\nu\nu}(\nu)}
\int_{-1}^\nu d\rho 
\frac{\partial{D_{p\rho}}}{\partial{p}}\textcircled{1}
\frac{\partial{}}{\partial{p}}\frac{\partial{}}{\partial{p}}
\frac{\partial{g}}{\partial{\rho}} 
-
\int_{-1}^\mu d\nu e^{-M(\nu)} \frac{1}{D_{\nu\nu}(\nu)}
\int_{-1}^\nu d\rho 
\left(\frac{\partial{}}{\partial{p}}\frac{\partial{D_{p\rho}}}{\partial{p}}
\right)\textcircled{1}
\frac{\partial{}}{\partial{p}}
\frac{\partial{g}}{\partial{\rho}} 
\nonumber\\
&&
-\frac{\partial{F}}{\partial{p}}
\int_{-1}^\mu d\nu e^{-M(\nu)} \frac{1}{D_{\nu\nu}(\nu)}
\int_{-1}^\nu d\rho 
\frac{\partial{}}{\partial{p}}\frac{\partial{}}{\partial{p}}
\left(\frac{1}{p^2}\frac{\partial{}}{\partial{p}}
\left(p^2
D_{pp}\right)\right)\textcircled{2} 
\nonumber\\
&&
-
\int_{-1}^\mu d\nu e^{-M(\nu)} \frac{1}{D_{\nu\nu}(\nu)}
\int_{-1}^\nu d\rho \frac{\partial{g}}{\partial{p}}
\frac{\partial{}}{\partial{p}}\frac{\partial{}}{\partial{p}}
\left(\frac{1}{p^2}\frac{\partial{}}{\partial{p}}
\left(p^2D_{pp}\right)\right)\textcircled{2} 
\nonumber\\
&&
-
2\int_{-1}^\mu d\nu e^{-M(\nu)} \frac{1}{D_{\nu\nu}(\nu)}
\int_{-1}^\nu d\rho 
\left(\frac{\partial{}}{\partial{p}}
\frac{\partial{g}}{\partial{p}}\right)
\frac{\partial{}}{\partial{p}}
\left(\frac{1}{p^2}\frac{\partial{}}{\partial{p}}
\left(p^2D_{pp}\right)\right)\textcircled{2} 
\nonumber\\
&&
-
\int_{-1}^\mu d\nu e^{-M(\nu)} \frac{1}{D_{\nu\nu}(\nu)}
\int_{-1}^\nu d\rho 
\left(\frac{\partial{}}{\partial{p}}
\frac{\partial{}}{\partial{p}}\frac{\partial{g}}{\partial{p}}
\right)
\left(\frac{1}{p^2}\frac{\partial{}}{\partial{p}}
\left(p^2D_{pp}\right)\right)\textcircled{2} 
\nonumber\\
&&
-
\int_{-1}^\mu d\nu e^{-M(\nu)} \frac{1}{D_{\nu\nu}(\nu)}
\int_{-1}^\nu d\rho 
D_{pp}\textcircled{2}
\frac{\partial{}}{\partial{p}}\frac{\partial{}}{\partial{p}}
\frac{\partial{}}{\partial{p}}\frac{\partial{g}}{\partial{p}}
-
\int_{-1}^\mu d\nu e^{-M(\nu)} \frac{1}{D_{\nu\nu}(\nu)}
\int_{-1}^\nu d\rho 
\frac{\partial{D_{pp}}}{\partial{p}}\textcircled{2}
\frac{\partial{}}{\partial{p}}
\frac{\partial{}}{\partial{p}}\frac{\partial{g}}{\partial{p}}
\nonumber\\
&&
-
\int_{-1}^\mu d\nu e^{-M(\nu)} \frac{1}{D_{\nu\nu}(\nu)}
\int_{-1}^\nu d\rho 
\frac{\partial{D_{pp}}}{\partial{p}}\textcircled{2}
\frac{\partial{}}{\partial{p}}
\frac{\partial{}}{\partial{p}}\frac{\partial{g}}{\partial{p}}
-
\int_{-1}^\mu d\nu e^{-M(\nu)} \frac{1}{D_{\nu\nu}(\nu)}
\int_{-1}^\nu d\rho 
\left(\frac{\partial{}}{\partial{p}}\frac{\partial{D_{pp}}}{\partial{p}}
\right)\textcircled{2}
\frac{\partial{}}{\partial{p}}\frac{\partial{g}}{\partial{p}}
\nonumber\\
&&+\frac{\partial{F}}{\partial{p}}\frac{1}{2}
\int_{-1}^\mu d\nu e^{-M(\nu)} \frac{1}{D_{\nu\nu}(\nu)}
\int_{-1}^1d\mu 
\frac{\partial{}}{\partial{p}}\frac{\partial{}}{\partial{p}}
\left(\frac{1}{p^2}\frac{\partial{}}{\partial{p}}
\left(p^2D_{pp}\right)\right)\textcircled{2}
\nonumber\\
&&
+\frac{1}{2}
\int_{-1}^\mu d\nu e^{-M(\nu)} \frac{1}{D_{\nu\nu}(\nu)}
\int_{-1}^1d\mu \frac{\partial{g}}{\partial{\mu}}
\frac{\partial{}}{\partial{p}}\frac{\partial{}}{\partial{p}}
\left(\frac{1}{p^2}\frac{\partial{}}{\partial{p}}
\left(p^2D_{p\mu}\right)\right)\textcircled{1}
\nonumber\\
&&
+\int_{-1}^\mu d\nu e^{-M(\nu)} \frac{1}{D_{\nu\nu}(\nu)}
\int_{-1}^1d\mu 
\left(\frac{\partial{}}{\partial{p}}\frac{\partial{g}}{\partial{\mu}}
\right)
\frac{\partial{}}{\partial{p}}
\left(\frac{1}{p^2}\frac{\partial{}}{\partial{p}}
\left(p^2D_{p\mu}\right)\right)\textcircled{1} 
\nonumber\\
&&
+\frac{1}{2}
\int_{-1}^\mu d\nu e^{-M(\nu)} \frac{1}{D_{\nu\nu}(\nu)}
\int_{-1}^1d\mu 
\left(\frac{\partial{}}{\partial{p}}
\frac{\partial{}}{\partial{p}}\frac{\partial{g}}{\partial{\mu}}
\right)
\left(\frac{1}{p^2}\frac{\partial{}}{\partial{p}}
\left(p^2D_{p\mu}\right)\right)\textcircled{1}
\nonumber\\
&&
+\frac{1}{2}
\int_{-1}^\mu d\nu e^{-M(\nu)} \frac{1}{D_{\nu\nu}(\nu)}
\int_{-1}^1d\mu 
D_{p\mu}\textcircled{1}
\frac{\partial{}}{\partial{p}}
\frac{\partial{}}{\partial{p}}
\frac{\partial{}}{\partial{p}}\frac{\partial{g}}{\partial{\mu}}
\nonumber\\
&&
+\int_{-1}^\mu d\nu e^{-M(\nu)} \frac{1}{D_{\nu\nu}(\nu)}
\int_{-1}^1d\mu 
\frac{\partial{D_{p\mu}}}{\partial{p}}\textcircled{1}
\frac{\partial{}}{\partial{p}}
\frac{\partial{}}{\partial{p}}\frac{\partial{g}}{\partial{\mu}}
+\frac{1}{2}
\int_{-1}^\mu d\nu e^{-M(\nu)} \frac{1}{D_{\nu\nu}(\nu)}
\int_{-1}^1d\mu 
\left(\frac{\partial{}}{\partial{p}}\frac{\partial{D_{p\mu}}}{\partial{p}}
\right)\textcircled{1}
\frac{\partial{}}{\partial{p}}\frac{\partial{g}}{\partial{\mu}}
\nonumber\\
&&
+\frac{1}{2}
\int_{-1}^\mu d\nu e^{-M(\nu)} \frac{1}{D_{\nu\nu}(\nu)}
\int_{-1}^1d\mu 
\frac{\partial{g}}{\partial{p}}
\frac{\partial{}}{\partial{p}}\frac{\partial{}}{\partial{p}}
\left(\frac{1}{p^2}\frac{\partial{}}{\partial{p}}
\left(p^2D_{pp}\right)\right)\textcircled{2}
\nonumber\\
&&
+
\int_{-1}^\mu d\nu e^{-M(\nu)} \frac{1}{D_{\nu\nu}(\nu)}
\int_{-1}^1d\mu 
\left(\frac{\partial{}}{\partial{p}}\frac{\partial{g}}{\partial{p}}
\right)
\frac{\partial{}}{\partial{p}}
\left(\frac{1}{p^2}\frac{\partial{}}{\partial{p}}
\left(p^2D_{pp}\right)\right)\textcircled{2} 
\nonumber\\
&&
+\frac{1}{2}
\int_{-1}^\mu d\nu e^{-M(\nu)} \frac{1}{D_{\nu\nu}(\nu)}
\int_{-1}^1d\mu 
\left(\frac{\partial{}}{\partial{p}}
\frac{\partial{}}{\partial{p}}\frac{\partial{g}}{\partial{p}}
\right)
\left(\frac{1}{p^2}\frac{\partial{}}{\partial{p}}
\left(p^2D_{pp}\right)\right)\textcircled{2} 
\nonumber\\
&&
+\frac{1}{2}
\int_{-1}^\mu d\nu e^{-M(\nu)} \frac{1}{D_{\nu\nu}(\nu)}
\int_{-1}^1d\mu 
D_{pp}\textcircled{2}
\frac{\partial{}}{\partial{p}}
\frac{\partial{}}{\partial{p}}
\frac{\partial{}}{\partial{p}}\frac{\partial{g}}{\partial{p}}
+
\int_{-1}^\mu d\nu e^{-M(\nu)} \frac{1}{D_{\nu\nu}(\nu)}
\int_{-1}^1d\mu 
\frac{\partial{D_{pp}}}{\partial{p}}\textcircled{2}
\frac{\partial{}}{\partial{p}}
\frac{\partial{}}{\partial{p}}\frac{\partial{g}}{\partial{p}}
\nonumber\\
&&
+\frac{1}{2}
\int_{-1}^\mu d\nu e^{-M(\nu)} \frac{1}{D_{\nu\nu}(\nu)}
\int_{-1}^1d\mu 
\left(\frac{\partial{}}{\partial{p}}\frac{\partial{D_{pp}}}{\partial{p}}
\right)\textcircled{2}
\frac{\partial{}}{\partial{p}}\frac{\partial{g}}{\partial{p}},
\label{R/pp-1}
\end{eqnarray}
and 
\begin{eqnarray}
\frac{\partial^2{R}}{\partial{\mu}\partial{p}}
\Longrightarrow&&
-\frac{\partial{F}}{\partial{p}} e^{-M(\mu)} \frac{\partial{}}{\partial{p}}
\frac{D_{\mu p}}{D_{\mu\mu}(\mu)} 
\textcircled{1}
-e^{-M(\mu)} \frac{D_{\mu p}}{D_{\mu\mu}(\mu)}
\textcircled{1} \frac{\partial{}}{\partial{p}}\frac{\partial{g}}{\partial{p}}
-e^{-M(\mu)} 
\frac{\partial{g}}{\partial{p}}
\frac{\partial{}}{\partial{p}}
\left(\frac{D_{\mu p}}{D_{\mu\mu}(\mu)}\right)\textcircled{1} 
\nonumber\\
&&
-\frac{e^{-M(\mu)}}{D_{\mu\mu}(\mu)}
\int_{-1}^\mu d\nu
\frac{\partial{g}}{\partial{\nu}}
\frac{\partial{}}{\partial{p}}
\left( \frac{1}{p^2}\frac{\partial{}}{\partial{p}}
\left(p^2D_{p\nu}\right)\right)\textcircled{1}
-\frac{e^{-M(\mu)}}{D_{\mu\mu}(\mu)}
\int_{-1}^\mu d\nu
\left(\frac{\partial{}}{\partial{p}}
\frac{\partial{g}}{\partial{\nu}}
\right)
\left( \frac{1}{p^2}\frac{\partial{}}{\partial{p}}
\left(p^2D_{p\nu}\right)\right)\textcircled{1}
\nonumber\\
&&
-\frac{e^{-M(\mu)}}{D_{\mu\mu}(\mu)}
\int_{-1}^\mu d\nu
D_{p\nu}\textcircled{1}
\frac{\partial{}}{\partial{p}}
\frac{\partial{}}{\partial{p}}\frac{\partial{g}}{\partial{\nu}}
-\frac{e^{-M(\nu)}}{D_{\nu\nu}(\nu)}
\int_{-1}^\nu d\rho
\frac{\partial{D_{p\rho}}}{\partial{p}}\textcircled{1}
\frac{\partial{}}{\partial{p}}\frac{\partial{g}}{\partial{\rho}}
\nonumber\\
&&
-\frac{\partial{F}}{\partial{p}}
\frac{e^{-M(\mu)}}{D_{\mu\mu}(\mu)}
\int_{-1}^\mu d\nu 
\frac{\partial{}}{\partial{p}}
\left( \frac{1}{p^2}\frac{\partial{}}{\partial{p}}
\left(p^2D_{pp}\right)\right)\textcircled{2}
-\frac{e^{-M(\mu)}}{D_{\mu\mu}(\mu)}
\int_{-1}^\mu d\nu 
\frac{\partial{g}}{\partial{p}}
\frac{\partial{}}{\partial{p}}
\left(\frac{1}{p^2}\frac{\partial{}}{\partial{p}}
\left(p^2D_{pp}\right)\right)\textcircled{2}
\nonumber\\
&&
-\frac{e^{-M(\mu)}}{D_{\mu\mu}(\mu)}
\int_{-1}^\mu d\nu 
\left(\frac{\partial{}}{\partial{p}}\frac{\partial{g}}{\partial{p}}
\right)
\frac{1}{p^2}\frac{\partial{}}{\partial{p}}
\left(p^2D_{pp}\right)\textcircled{2}
-\frac{e^{-M(\mu)}}{D_{\mu\mu}(\mu)}
\int_{-1}^\mu d\nu 
D_{pp}\textcircled{2}
\frac{\partial{}}{\partial{p}}
\frac{\partial{}}{\partial{p}}\frac{\partial{g}}{\partial{p}}
\nonumber\\
&&
-\frac{e^{-M(\mu)}}{D_{\mu\mu}(\mu)}
\int_{-1}^\mu d\nu 
\frac{\partial{D_{pp}}}{\partial{p}}\textcircled{2}
\frac{\partial{}}{\partial{p}}\frac{\partial{g}}{\partial{p}}
+\frac{\partial{F}}{\partial{p}}
\frac{e^{-M(\mu)}}{D_{\mu\mu}(\mu)} 
\frac{1}{2}\int_{-1}^1d\mu 
\frac{\partial{}}{\partial{p}}
\left(\frac{1}{p^2}\frac{\partial{}}{\partial{p}}
\left(p^2D_{pp}\right)\right)\textcircled{2}
\nonumber\\
&&
+\frac{e^{-M(\mu)}}{D_{\mu\mu}(\mu)}
\frac{1}{2}\int_{-1}^1d\mu \frac{\partial{g}}{\partial{\mu}}
\frac{\partial{}}{\partial{p}}
\left(\frac{1}{p^2}\frac{\partial{}}{\partial{p}}
\left(p^2D_{p\mu}\right)\right)\textcircled{1}
+\frac{e^{-M(\mu)}}{D_{\mu\mu}(\mu)}
\frac{1}{2}\int_{-1}^1d\mu 
\left(\frac{\partial{}}{\partial{p}}\frac{\partial{g}}{\partial{\mu}}
\right)
\left(\frac{1}{p^2}\frac{\partial{}}{\partial{p}}
\left(p^2D_{p\mu}\right)\right)\textcircled{1}
\nonumber\\
&&
+\frac{e^{-M(\mu)}}{D_{\mu\mu}(\mu)}
\frac{1}{2}\int_{-1}^1d\mu 
D_{p\mu}\textcircled{1}
\frac{\partial{}}{\partial{p}}
\frac{\partial{}}{\partial{p}}\frac{\partial{g}}{\partial{\mu}}
+\frac{e^{-M(\mu)}}{D_{\mu\mu}(\mu)}
\frac{1}{2}\int_{-1}^1d\mu 
\frac{\partial{D_{p\mu}}}{\partial{p}}\textcircled{1}
\frac{\partial{}}{\partial{p}}\frac{\partial{g}}{\partial{\mu}}
\nonumber\\
&&
+\frac{e^{-M(\mu)}}{D_{\mu\mu}(\mu)} 
\frac{1}{2}\int_{-1}^1d\mu 
\frac{\partial{g}}{\partial{p}}
\frac{\partial{}}{\partial{p}}
\left(\frac{1}{p^2}\frac{\partial{}}{\partial{p}}
\left(p^2D_{pp}\right)\right)\textcircled{2}
+\frac{e^{-M(\mu)} }{D_{\mu\mu}(\mu)} 
\frac{1}{2}\int_{-1}^1d\mu 
\left(\frac{\partial{}}{\partial{p}}\frac{\partial{g}}{\partial{p}}
\right)
\left(\frac{1}{p^2}\frac{\partial{}}{\partial{p}}
\left(p^2D_{pp}\right)\right)\textcircled{2}
\nonumber\\
&&
+\frac{e^{-M(\mu)}}{D_{\mu\mu}(\mu)} 
\frac{1}{2}\int_{-1}^1d\mu 
D_{pp}\textcircled{2}
\frac{\partial{}}{\partial{p}}
\frac{\partial{}}{\partial{p}}\frac{\partial{g}}{\partial{p}}
+\frac{e^{-M(\mu)}}{D_{\mu\mu}(\mu)} 
\frac{1}{2}\int_{-1}^1d\mu 
\frac{\partial{D_{pp}}}{\partial{p}}\textcircled{2}
\frac{\partial{}}{\partial{p}}\frac{\partial{g}}{\partial{p}}. 
\label{R/pmu-1}
\end{eqnarray}

\section{The formula of $\MakeLowercase{\partial{g}/\partial{\mu}}$}
\label{The formula of g/mu}


\section{The formula of $\MakeLowercase{\partial{g}/\partial{p}}$}
\label{The formula of g/p}
\begin{eqnarray}
\frac{\partial{g}}{\partial{p}}
&&\Longrightarrow
+\frac{\partial{F}}{\partial{p}}
\left(\frac{2e^{M(\mu)}}{\int_{-1}^{1}
	d\mu e^{M(\mu) }}-1\right)\nonumber\\
&&-\frac{\partial{F}}{\partial{p}}
e^{M(\mu)}\int_{-1}^\mu d\nu e^{-M(\nu)} \frac{\partial{}}{\partial{p}}
\frac{D_{\nu p}}{D_{\nu\nu}(\nu)} 
\textcircled{1}
\nonumber\\
&&
-\frac{\partial{F}}{\partial{p}}
e^{M(\mu)}\int_{-1}^\mu d\nu e^{-M(\nu)} 
\left(\frac{2e^{M(\nu)}}{\int_{-1}^{1}
	d\mu e^{M(\mu) }}-1\right)
\frac{\partial{}}{\partial{p}}
\left(\frac{D_{\nu p}}{D_{\nu\nu}(\nu)}\right)\textcircled{1} 
\nonumber\\
&&
-\frac{\partial{F}}{\partial{p}}e^{M(\mu)}\int_{-1}^\mu d\nu
\frac{e^{-M(\nu)}}{D_{\nu\nu}(\nu)}
\int_{-1}^\nu d\rho
\frac{2\frac{\partial{e^{M(\rho)}}}{\partial{\rho}}}{\int_{-1}^{1}
d\mu e^{M(\mu) }}
\left( \frac{1}{p^2}\frac{\partial{}}{\partial{p}}
\left(p^2D_{p\rho}\right)\right)\textcircled{1}
\nonumber\\
&&
-\frac{\partial{F}}{\partial{p}}
e^{M(\mu)}\int_{-1}^\mu d\nu \frac{e^{-M(\nu)}}{D_{\nu\nu}(\nu)}
\int_{-1}^\nu d\rho
\frac{\partial{D_{p\rho}}}{\partial{p}}\textcircled{1}
\frac{2\frac{\partial{e^{M(\rho)}}}{\partial{\rho}}}{\int_{-1}^{1}
	d\mu e^{M(\mu) }}
\nonumber\\
&&
+\frac{\partial{F}}{\partial{p}}
e^{M(\mu)}\int_{-1}^\mu d\nu \frac{e^{-M(\nu)}}{D_{\nu\nu}(\nu)}
\frac{1}{2}\int_{-1}^1d\mu 
\frac{2\frac{\partial{e^{M(\mu)}}}{\partial{\mu}}}{\int_{-1}^{1}
	d\mu e^{M(\mu) }}
\left(\frac{1}{p^2}\frac{\partial{}}{\partial{p}}
\left(p^2D_{p\mu}\right)\right)\textcircled{1}
\nonumber\\
&&
+\frac{\partial{F}}{\partial{p}}
e^{M(\mu)}\int_{-1}^\mu d\nu \frac{e^{-M(\nu)}}{D_{\nu\nu}(\nu)}
\frac{1}{2}\int_{-1}^1d\mu 
\frac{\partial{D_{p\mu}}}{\partial{p}}\textcircled{1}
\frac{2\frac{\partial{e^{M(\mu)}}}{\partial{\mu}}}{\int_{-1}^{1}
	d\mu e^{M(\mu) }}
\nonumber\\
&&+\frac{\partial{F}}{\partial{p}}
\frac{1}
{\int_{-1}^{1}d\mu
	e^{M(\mu) }}e^{M(\mu)}
\int_{-1}^{1}d\mu
e^{M(\mu)}
\int_{-1}^\mu d\nu e^{-M(\nu)} \frac{\partial{}}{\partial{p}}
\frac{D_{\nu p}}{D_{\nu\nu}(\nu)} 
\textcircled{1}
\nonumber\\
&&
+\frac{\partial{F}}{\partial{p}}\frac{1}
{\int_{-1}^{1}d\mu
	e^{M(\mu) }}e^{M(\mu)}
\int_{-1}^{1}d\mu
e^{M(\mu)}\int_{-1}^\mu d\nu e^{-M(\nu)} 
\left(\frac{2e^{M(\nu)}}{\int_{-1}^{1}
	d\mu e^{M(\mu) }}-1\right)
\frac{\partial{}}{\partial{p}}
\left(\frac{D_{\nu p}}{D_{\nu\nu}(\nu)}\right)\textcircled{1} 
\nonumber\\
&&
+\frac{\partial{F}}{\partial{p}}\frac{1}
{\int_{-1}^{1}d\mu
	e^{M(\mu) }}e^{M(\mu)}
\int_{-1}^{1}d\mu
e^{M(\mu)}
\int_{-1}^\mu d\nu \frac{e^{-M(\nu)}}{D_{\nu\nu}(\nu)}
\int_{-1}^\nu d\rho
\frac{2\frac{\partial{e^{M(\rho)}}}{\partial{\rho}}}{\int_{-1}^{1}
	d\mu e^{M(\mu) }}
\left( \frac{1}{p^2}\frac{\partial{}}{\partial{p}}
\left(p^2D_{p\rho}\right)\right)\textcircled{1}
\nonumber\\
&&
+\frac{\partial{F}}{\partial{p}}\frac{1}
{\int_{-1}^{1}d\mu
	e^{M(\mu) }}e^{M(\mu)}
\int_{-1}^{1}d\mu
e^{M(\mu)}
\int_{-1}^\mu d\mu \frac{e^{-M(\mu)}}{D_{\mu\mu}(\mu)}
\int_{-1}^\mu d\mu
\frac{\partial{D_{p\mu}}}{\partial{p}}\textcircled{1}
\frac{2\frac{\partial{e^{M(\mu)}}}{\partial{\mu}}}{\int_{-1}^{1}
	d\mu e^{M(\mu) }}
\nonumber\\
&&
-\frac{\partial{F}}{\partial{p}}\frac{1}
{\int_{-1}^{1}d\mu
	e^{M(\mu) }}e^{M(\mu)}
\int_{-1}^{1}d\mu
e^{M(\mu)}
\int_{-1}^\mu d\nu \frac{e^{-M(\nu)}}{D_{\nu\nu}(\nu)}
\frac{1}{2}\int_{-1}^1d\mu 
\frac{2\frac{\partial{e^{M(\mu)}}}{\partial{\mu}}}{\int_{-1}^{1}
	d\mu e^{M(\mu) }}
\left(\frac{1}{p^2}\frac{\partial{}}{\partial{p}}
\left(p^2D_{p\mu}\right)\right)\textcircled{1}
\nonumber\\
&&
-\frac{\partial{F}}{\partial{p}}\frac{1}
{\int_{-1}^{1}d\mu
	e^{M(\mu) }}e^{M(\mu)}
\int_{-1}^{1}d\mu
e^{M(\mu)}
\int_{-1}^\mu d\nu \frac{e^{-M(\nu)}}{D_{\nu\nu}(\nu)}
\frac{1}{2}\int_{-1}^1d\mu 
\frac{\partial{D_{p\mu}}}{\partial{p}}\textcircled{1}
\frac{2\frac{\partial{e^{M(\mu)}}}{\partial{\mu}}}{\int_{-1}^{1}d\mu e^{M(\mu) }}
\label{g/p-10}
\end{eqnarray}

\section{Evaluating $A$}
\label{Evaluating A}

The momentum diffusion coefficient with the focusing effect
is shown as
\begin{eqnarray}
A_4(L)=A+A'+\mathcal{M}_4(\epsilon,\xi)
\end{eqnarray}
with
\begin{eqnarray}
\mathcal{M}_4(\epsilon,\xi)=\sum_{n=1}^{253}T_n(\epsilon,\xi)
\end{eqnarray}
Here, $A$
is the momentum diffusion coefficients exact up to  
second order in $\epsilon$ for constant field,  
and $A'$ is the third- and fourth-order coefficient for uniform field. 
$\mathcal{M}_4(\epsilon,\xi)$ is the coefficient of the additional 
SOMT caused by focused field. 
In what follows, we evaulate all of the terms of $\mathcal{M}_4(\epsilon,\xi)$. In this process, we set the terms 
higher than fourth order of $\epsilon$ as zero. For simplification,
the third- and fourth-order terms for constant
field are also ignored, which are not the research topic 
of this paper.   
\begin{eqnarray}
&&T_1
=-\frac{2}{45}\epsilon^2 p^2 D_1 \xi^2; 
T_2
=-2\frac{D_2^2}{D_1}\epsilon^2 p^2 
\left(\frac{1}{3}\xi
+\frac{1}{5}\xi^2\right);
T_3
=-\frac{152}{135}p^2\epsilon^3D_2\xi;
T_4=\frac{4}{9}\epsilon^3 p^2D_2\xi
\int_{-1}^1d\mu
\frac{2-3\mu+\mu^3}{1-\mu^2};
\nonumber\\
&&T_5=\frac{4}{15}
\frac{D_2^2}{D_1}\epsilon^2 p^2 \xi^2;
T_6=\frac{2}{5}\frac{ D_2^3}{D_1^2}
\epsilon^3 p^2 \xi;
T_7=T_8=T_9=T_{10}=T_{11}=0;
T_{12}=-\frac{2}{3} \epsilon^3 p^2 \frac{D_2^3}{D_1^2}\xi;
\nonumber\\
&&
T_{13}=T_{14}=T_{15}=T_{16}=T_{17}=T_{18}=T_{19}=0=T_{20}
=T_{21}=T_{22}=T_{23}=T_{24}=T_{25}=0;
\nonumber\\
&&T_{26}=\frac{38}{45}\epsilon^3 p^2 
\frac{D_2^3}{D_1^2} \xi;
T_{27}=T_{28}=T_{29}=0;
T_{30}=-\frac{38}{135}\epsilon^2 p^2  
\frac{D_2^2}{D_1}\xi^2;
T_{31}=T_{32}=T_{33}=T_{34}=T_{35}=T_{36}
\nonumber\\
&&=T_{37}=T_{38}=T_{39}=T_{40}=T_{41}=T_{42}=0;
T_{43}=\frac{38}{135}\epsilon^3 p^2 
\frac{D_2^3}{D_1^2}\xi;
T_{44}=T_{45}=T_{46}=T_{47}=T_{48}=T_{49}=
\nonumber\\
&&=T_{50}=T_{51}=T_{52}=T_{53}=T_{54}=T_{55}=
T_{56}=T_{57}=0;
T_{58}=-
\frac{1}{3}\epsilon^3 p^2 \frac{D_2^3}{D_1^2}\xi
\int_{-1}^1d\mu\frac{2-3\mu+\mu^3}{1-\mu^2};
\nonumber\\
&&T_{59}=T_{60}=T_{61}=0;
T_{62}=\frac{1}{9}\epsilon^2 p^2 \frac{D_2^2}{D_1} \xi^2
\int_{-1}^1d\mu\frac{2-3\mu+\mu^3}{1-\mu^2};
T_{63}=T_{64}=T_{65}=T_{66}=T_{67}=T_{68}=T_{69}
\nonumber\\
&&=T_{70}=T_{71}=T_{72}=T_{73}=T_{74}=0;
T_{75}=-\frac{1}{9}\epsilon^3 p^2 
\frac{D_2^3}{D_1^2} \xi
\int_{-1}^1d\mu\frac{2-3\mu+\mu^3}{1-\mu^2};
T_{76}=T_{77}=T_{78}=T_{79}
\nonumber\\
&&=T_{80}=T_{81}=0;
T_{82}=\frac{1}{3}\epsilon^2 p^2 
\frac{D_2^2}{D_1}
\left(\xi^2\frac{10}{3}+2\xi\right);
T_{83}=\frac{40}{27}\epsilon^3 p^2 D_2 \xi;
T_{84}=-\frac{8}{9}\epsilon^3 p^2 D_2 \xi
\int_{-1}^{1}d\mu
\frac{1-\mu}{1-\mu^2};
\nonumber\\
&&T_{85}=-\frac{2}{9}
\epsilon^2 p^2 \frac{D_2^2}{D_1}\xi^2;
T_{86}=-\frac{4}{9}\epsilon^3 p^2 
\frac{D_2^3}{D_1^2}\xi;
T_{87}=T_{88}=T_{89}=T_{90}=T_{91}=0;
T_{92}=\frac{2}{3}
\epsilon^3 p^2\frac{D_2^3}{D_1^2}\xi;
\nonumber\\
&&T_{93}=T_{94}=T_{95}=T_{96}=T_{97}=T_{98}=
T_{99}=T_{100}=T_{101}=T_{102}=T_{103}=T_{104}=T_{105}=0
\nonumber\\
&&T_{106}=-\frac{10}{9}\epsilon^3 p^2
\frac{D_2^3}{D_1^2}\xi;
T_{107}=T_{108}=T_{109}=0;
T_{110}=\frac{20}{27}
\epsilon^2 p^2 \frac{ D_2^2}{D_1}\xi^2;
T_{111}=T_{112}=T_{113}=T_{114}
=T_{115}
\nonumber\\
&&
=T_{116}=T_{117}=T_{118}=T_{119}=T_{120}=T_{121}=T_{122}=0;
T_{123}=
-\frac{10}{27}
\epsilon^3 p^2 \frac{D_2^3}{D_1^2}\xi;
T_{124}=T_{125}=T_{126}=T_{127}
\nonumber\\
&&=T_{128}=T_{129}=T_{130}=T_{131}=T_{132}=T_{133}
=T_{134}=T_{135}=T_{136}=T_{137}=0;
T_{138}=\frac{2}{3}
\epsilon^3 p^2 \frac{D_2^3}{D_1^2}\xi
\int_{-1}^{1}d\mu
\frac{1-\mu}{1-\mu^2}
\nonumber\\
&&T_{139}=T_{140}=T_{141}=0;
T_{142}=-\frac{2}{9}\epsilon^2 p^2 \frac{D_2^2}{D_1} \xi^2
\int_{-1}^{1}d\mu \frac{1-\mu}{1-\mu^2};
T_{143}=T_{144}=T_{145}=T_{146}=T_{147}=T_{148}=
T_{149}
\nonumber\\
&&=T_{150}=T_{151}=T_{152}=T_{153}=T_{154}=0;
T_{155}=\frac{2}{9}\epsilon^3 p^2 \frac{D_2^3}{D_1^2}\xi
\int_{-1}^{1}d\mu \frac{1-\mu}{1-\mu^2};
T_{156}=T_{157}=T_{158}=T_{159}=T_{160}=0
\nonumber\\
&&=T_{161}=0;
T_{162}=
-\frac{1}{2}\int_{-1}^1d\mu
D_{p\mu}
\frac{D_{\mu p}}{D_{\mu\mu}};
T_{163}=-\frac{8}{3}\epsilon^3 p^2 D_2;
T_{164}=\frac{8}{3}\epsilon^3 p^2 D_2;
T_{165}=\frac{2}{45}
\epsilon^2 p^2 \frac{D_2^2}{D_1} \xi^2;
\nonumber\\
&&T_{166}=\frac{2}{15}\epsilon^3 p^2 \frac{D_2^3}{D_1^2}
(3\xi+5);
T_{167}=-\frac{4}{15}
\epsilon^3 p^2 \frac{D_2^3}{D_1^2}\xi;
T_{168}=\frac{38}{45}
\epsilon^3 p^2 
\frac{D_2^3}{D_1^2}\xi;
T_{169}=\frac{38}{135}
\epsilon^3 p^2 \frac{D_2^3}{D_1^2}\xi;
\nonumber\\
&&T_{170}=-\frac{1}{3}
\epsilon^3 p^2\frac{D_2^3}{D_1^2}\xi
\int_{-1}^1d\mu\frac{2-3\mu+\mu^3}{1-\mu^2};
T_{171}=-\frac{1}{9}
\epsilon^3 p^2 \frac{D_2^3}{D_1^2}\xi
\int_{-1}^1d\mu\frac{2-3\mu+\mu^3}{1-\mu^2};
T_{172}=-\frac{1}{3}
\epsilon^3 p^2 \frac{D_2^3}{D_1^2}
\Bigg(2+\frac{10}{3}\xi\Bigg)
\nonumber\\
&&T_{173}=\frac{2}{9}\epsilon^3 p^2  
\frac{D_2^3}{D_1^2}\xi;
T_{174}=-\frac{10}{9}\epsilon^3 p^2 
\frac{D_2^3}{D_1^2}\xi;
T_{175}=-\frac{10}{27}
\epsilon^3 p^2 \frac{D_2^3}{D_1^2}\xi;
T_{176}=\frac{2}{3}
\epsilon^3 p^2 \frac{D_2^3}{D_1^2}\xi
\int_{-1}^{1}d\mu \frac{1-\mu}{1-\mu^2};
\nonumber\\
&&T_{177}=
\frac{2}{9}
\epsilon^3 p^2 \frac{D_2^3}{D_1^2}\xi
\int_{-1}^{1}d\mu \frac{1-\mu}{1-\mu^2};
T_{178}=
\frac{8}{5} \epsilon^3 p^2 
\frac{D_2^3}{D_1^2}\xi;
T_{179}=T_{180}=T_{181}=0;
T_{182}=-2\epsilon^3 p^2 \frac{D_2^3}{D_1^2} \xi;
\nonumber\\
&&T_{183}=T_{184}=T_{185}=0;
T_{186}=2\epsilon^3 p^2
\frac{D_2^3}{D_1^2}; 
T_{187}=-\frac{2}{5} \epsilon^3 p^2 
\frac{D_2^3}{D_1^2}\xi;
T_{188}=\frac{6}{5}
\epsilon^3 p^2\frac{ D_2^3}{D_1^2} \xi;
T_{189}=-2\epsilon^3p^2 \frac{D_2^3}{D_1^2} \xi;
\nonumber\\
&&T_{190}=
-\frac{1}{2}
\epsilon^2 p^2\frac{D_2^2}{D_1}  
\left(\xi \frac{4}{3}
+\xi^2\frac{16}{15}\right);
T_{191}=
\frac{8}{15}
\epsilon^3 p^2 \frac{D_2^3}{D_1^2} \xi;
T_{192}=T_{193}=T_{194}=T_{195}=
T_{196}=0;
T_{197}=
-\frac{2}{3}
\epsilon^3 p^2 \frac{D_2^3}{D_1^2}\xi
\nonumber\\
&&T_{198}=T_{199}=T_{200}=T_{201}=
T_{202}=0;
T_{203}=\frac{2}{3}
\frac{D_2^3}{D_1^2}\epsilon^3 p^2; 
T_{204}=
-\frac{2}{15}
\epsilon^3 p^2 \frac{D_2^3}{D_1^2}\xi;
T_{205}=
\frac{6}{5}
\epsilon^3 p^2 \frac{D_2^3}{D_1^2}\xi;
\nonumber\\
&&
T_{206}=
\frac{2}{5}
\epsilon^3 p^2 \frac{D_2^3}{D_1^2} \xi;
T_{207}=
-2\epsilon^3p^2 \frac{D_2^3}{D_1^2}\xi;
T_{208}=
-\frac{2}{3}
\epsilon^3 p^2 \frac{D_2^3}{D_1^2}\xi;
T_{209}=\frac{8}{15}
\epsilon^3 p^2 D_2 \xi;
T_{210}=-2\epsilon^3 p^2 \frac{D_2^3}{D_1^2} \xi;
\nonumber\\
&&T_{211}=T_{212}=T_{213}=0;
T_{214}
=2\epsilon^3 p^2 \frac{D_2^3}{D_1^2} \xi;
T_{215}=T_{216}=T_{217}=0;
T_{218}=-2\epsilon^3 p^2 \frac{D_2^3}{D_1^2};
T_{219}=0;
\nonumber\\
&&T_{220}
=-2\epsilon^3 p^2 \frac{D_2^3}{D_1^2}\xi;
T_{221}=2\epsilon^3 p^2 \frac{D_2^3}{D_1^2} \xi;
T_{222}=\frac{2}{3}
\epsilon^2 p^2\frac{D_2^2}{D_1}  
(\xi+\xi^2);
T_{223}=-\frac{2}{3}
\epsilon^3 p^2 \frac{D_2^3}{D_1^2}\xi;
T_{224}=T_{225}
\nonumber\\
&&=T_{226}=T_{227}=T_{228}=0;
T_{229}=\frac{2}{3}
\epsilon^3 p^2 \frac{D_2^3}{D_1^2} \xi;
T_{230}=T_{231}=T_{232}=T_{233}=T_{234}=0;
T_{235}=
-\frac{2}{3}
\epsilon^3 p^2 \frac{D_2^3}{D_1^2}
\nonumber\\
&&T_{236}=0;
T_{237}=-2\epsilon^3 p^2 \frac{D_2^3}{D_1^2} \xi;
T_{238}=
-\frac{2}{3}  
\epsilon^3 p^2 \frac{D_2^3}{D_1^2} \xi;
T_{239}=2\epsilon^3 p^2 \frac{D_2^3}{D_1^2} \xi; 
T_{240}=
\frac{2}{3}
\epsilon^3 p^2 \frac{D_2^3}{D_1^2}\xi;
T_{241}=0;
\nonumber\\
&&T_{242}=-2\epsilon^3 p^2 D_2
\left(\frac{1}{3}+\xi\frac{1}{5}\right);
T_{243}=\frac{4}{15}
\epsilon^3 p^2 D_2\xi;
T_{244}=
-\frac{38}{45}
\epsilon^3 p^2 D_2\xi;
T_{245}
=-\frac{38}{135}
\epsilon^3 p^2 D_2 \xi;
\nonumber\\
&&T_{246}=
\frac{1}{3}\epsilon^3 p^2 D_2 \xi
\int_{-1}^1d\mu\frac{2-3\mu+\mu^3}{1-\mu^2};
T_{247}=
\frac{1}{9}
\epsilon^3 p^2 D_2\xi
\int_{-1}^1d\mu\frac{2-3\mu+\mu^3}{1-\mu^2};
T_{248}=\frac{2}{9}
\epsilon^3 p^2 D_2\left(1+3\xi\right);
\nonumber\\
&&T_{249}=-\frac{2}{9}
\epsilon^3 p^2 D_2\xi;
T_{250}=
\frac{10}{9}\epsilon^3 p^2 D_2\xi;
T_{251}=
\frac{10}{27}
\epsilon^3 p^2 D_2\xi;
T_{252}=
-\frac{2}{3}
\epsilon^3 p^2 D_2 \xi
\int_{-1}^{1}d\mu \frac{1-\mu}{1-\mu^2};
\nonumber\\
&&T_{253}=
-\frac{2}{9}
\epsilon^3 p^2 D_2\xi
\int_{-1}^{1}d\mu \frac{1-\mu}{1-\mu^2}.
\nonumber
\end{eqnarray}

\end{appendices}

\clearpage
\begin{table}[ht]
\begin{center}
\caption{The modifying term sign}
\label{The modifying term sign}
\begin{tabular}{|l|l|l|l|}
\hline
\multirow{3}{*}{Diverging field}   & \multirow{3}{*}{$\xi>0$} & $H_C=1$ & $\mathcal{M}_4<0$ \\
\cline{3-4} 
&                                & $H_C=0$ & $\mathcal{M}_4<0$ \\
\cline{3-4}
&                                & $H_C=-1$ & $\mathcal{M}_4>0$ \\
\hline
\multirow{3}{*}{Diverging field}   & \multirow{3}{*}{$\xi<0$} & $H_C=1$ & $\mathcal{M}_4>0$ \\
\cline{3-4} 
&                                & $H_C=0$ & $\mathcal{M}_4<0$ \\
\cline{3-4}
&                                & $H_C=-1$ & $\mathcal{M}_4<0$ \\
\hline
		\end{tabular}
	\end{center}
\end{table}


\begin{thebibliography}{}
\bibitem[Armstrong et al(2012)]{ArmstrongEA2012}Armstrong, C. K., 
Litvinenko, Y. E., \& Wibberenz, G. 2012, \apj, 757, 165
\bibitem[Beeck \& Wibberenz(1986)]{BeeckEA1986}Beeck, J., \& Wibberenz, G. 1986, \apj, 311, 437
\bibitem[Bian et al. (2017)]{BianEA2017}Bian, N. H., Emslie, A. G. \& Kontar, E. P. 2017, \apj,  835, 262
\bibitem[Bieber \& Burger(1990)]{BieberEA1990}Bieber, J. W., \& Burger, R. A. 1990, \apj, 348, 597
\bibitem[Chandrasekher(1943)]{Chandrasekher1943}Chandrasekher, S. 1943, \rmp, 15, 1
\bibitem[del-Castillo-Negrete et al. (2004)]
{NegreteEA2004}del-Castillo-Negrete, D.,
Carreras, B. A. \& Lynch, V. E. 2004, Physics of Plasmas, 11, 3854
\bibitem[del-Castillo-Negrete et al. (2005)]{NegreteEA2005}del-Castillo-Negrete, D.,
Carreras, B. A. \& Lynch, V. E. 2005, \prl, 
94, 065003
\bibitem[Dung \& Schlickeiser(1990)]{DungEA1990}Dung, R., \& Schlickeiser, R. 1990, \apss, 237, 504
\bibitem[Earl(1976)]{Earl1976}Earl, J. A. 1976, \apj, 205, 900
\bibitem[Earl(1981)]{Earl1981}Earl, J. A. 1981, \apj, 251, 739
\bibitem[Giacalone(2013)]{Giacalone2013}Giacalone, J. \ssr, 176, 73
\bibitem[Gombosi et al. (1993)]{GombosiEA1993}Gombosi, T. I., Jokipii, J. R.,
Kota,J., Lorencz, K. \& Williams, L. L. 1993, \apj, 403, 377
\bibitem[He \& Schlickeiser(2014)]{HeEA2014}He, H.-Q., \& Schlickeiser, R. 2014, \apj, 792, 85
\bibitem[Isichenko(1992)]{Icichenko1992}Isichenko, M. B. 1992, \rmp, 64, 961
\bibitem[Jokipii(1966)]{Jokipii1966}Jokipii, J. R. 1966, \apj, 146, 480
\bibitem[K\'ota(2000)]{Kota2000}K\'ota, J. 2000, \jgr, 105, 2403
\bibitem[Kulsrud(1979)]{Kulsrud1979}Kulsrud, R. 1979, AIP Conference Proceedings, 56, 13
\bibitem[Kunstmann(1979)]{Kunstmann1979}Kunstmann, J. E. 1979, \apj, 229, 812
\bibitem[Lasuik et al. (2017)]{LasuikEA2017}Lasuik, J., Fiege, D. J., \& Shalchi, A. 2017, Adv. Space Res., 59, 722
\bibitem[Lasuik \& Shalchi(2019)]{LasuikEA2019}Lasuik, J., \& Shalchi, A. 2019, \mnras,485, 1635
\bibitem[Lee et al. (2012)]{LeeEA2012}Lee, M., Mewaldt, R. A., \& Giacalone, J. 2012, \ssr, 173, 247
\bibitem[Lefa et al. (2011)]{LefaEA2011}Lefa, E., Rieger, F. M., \& Aharonian, F. 2011, \apj, 740, 64
\bibitem[Litvinenko(2012a)]{Litvinenko2012a}Litvinenko, Y. E. 2012a, \apj, 752, 16
\bibitem[Litvinenko(2012b)]{Litvinenko2012b}Litvinenko, Y. E. 2012b, \apj, 745, 62
\bibitem[Litvinenko \& Noble(2013)]{LitvinenkoEA2013}Litvinenko, Y. E., Noble, P. L. 2013, \apj, 765, 31
\bibitem[Litvinenko \& Schlickeiser(2011)]{LitvinenkoEA2011}Litvinenko, Y. E., \& Schlickeiser, R. A. 2011, \apj, 732, L31
\bibitem[Malkov(2017)]{Malkov2017}Malkov, M. A. 2017, \prd, 95, 023007
\bibitem[Malkov \& Sagdeev (2015)]{MalkovEA2015}Malkov, M. A., \& Sagdeev, R. Z. 2015, \apj, 808, 157
\bibitem[Matthaeus et al.(2003)]{MatthaeusEA2003}Matthaeus, W. H., Qin, G., Bieber, J. W., \& Zank, G. P. 2003, \apj, 590, L53
\bibitem[Mertsch \& Sarkar(2011)]{MertschEA2011}Mertsch, P., \& Sarkar, S. 2011, \prl, 107, 091101
\bibitem[O'Sullivan et al. (2009)]{O'SullivanEA2009}O'Sullivan, S., Reville, B., \& Taylor, A. M. 2009, \mnras, 400, 248
\bibitem[Parker(1958)]{Parker1958}Parker, E. N. 1958, \apj, 128, 664
\bibitem[Parker(1965)]{Parker1965}Parker, E. N. 1965, Planet. Space Sci., 13, 9
\bibitem[Petrosian(2012)]{Petrosian2012}Petrosian, V. 2012, \ssr, 173, 535
\bibitem[Qin(2007)]{Qin2007}Qin, G. 2007, \apj, 656, 217
\bibitem[Qin \& Zhang(2014)]{QinEA2014}Qin, G., \& Zhang, L.-H. 2014, \apj, 787, 1
\bibitem[Roelof(1969)]{Roelof1969}Roelof, E. C. 1969, in Lectures in High Energy Astrophysics, eds. H. \"Ogelman, J. R. Wayland (NASA SP-199; Washington, DC: NASA), 111 
\bibitem[Ruffolo(1995)]{Ruffolo1995}Ruffolo, D. 1995, \apj, 442, 861
\bibitem[Ruffolo et al. (1998)]{RuffoloEA1998}Ruffolo, D., Khumlumlert, T., \& Youngdee, W. 1998, \jgr, 103, 20591
\bibitem[Saiz et al.(2008)]{SaizEA2008}Saiz, A., Ruffolo, D., Bieber, J. W., Evenson, P. A., \& Pyle, R. 2003, \apj, 672, 650
\bibitem[Schlickeiser(2002)]{Schlickeiser2002}Schlickeiser, R. 2002, Cosmic Ray Astrophysics (Berlin: Springer)
\bibitem[Schlickeiser(1989a)]{Schlickeiser1989a}Schlickeiser, R. 1989a, \apj, 336, 243
\bibitem[Schlickeiser(1989b)]{Schlickeiser1989b}Schlickeiser, R. 1989b, \apj, 336, 264
\bibitem[Schlickeiser(2011)]{Schlickeiser2011}Schlickeiser, R. 2011, \apj, 732, 96
\bibitem[Schlickeiser et al(2007)]{SchlickeiserEA2007}Schlickeiser, R., Dohle, U., Tautz, R.C., \& Shalchi, A. 2007, \apj, 661, 185
\bibitem[Schlickeiser \& Jenko(2010)]{SchlickeiserEA2010}Schlickeiser, R., \& Jenko, F. 2010, J. Plasma Physics, 76, 317
\bibitem[Schlickeiser \& Shalchi(2008)]{SchlickeiserEA2008}Schlickeiser, R., \& Shalchi, A. 2008, \apj, 686, 292
\bibitem[Shalchi(2006)]{Shalchi2006}Shalchi, A. 2006, A\&A, 453, L43
\bibitem[Shalchi(2009a)]{Shalchi2009a}Shalchi, A. 2009a, J. Phys. G:Nucl. Part. Phys., 36, 025202
\bibitem[Shalchi(2009b)]{Shalchi2009b}Shalchi, A. 2009b, Nonlinear Cosmic Ray Diffusion Theories, Astrophysics and Space Science Library, Vol. 362 (Berlin: Springer)
\bibitem[Shalchi(2010)]{Shalchi2010}Shalchi, A. 2010, ApJL, 720, L127
\bibitem[Shalchi(2011)]{Shalchi2011}Shalchi, A. 2011, \apj, 728, 113
\bibitem[Shalchi(2017)]{Shalchi2017}Shalchi, A. 2017, Physics of Plasmas, 24, 050702
\bibitem[Shalchi(2020)]{Shalchi2020}Shalchi, A. 2020, \ssr, 216,23
\bibitem[Shalchi \& Danos(2013)]{ShalchiEA2013}Shalchi, A., \& Danos, R. J. 2013, \apj, 765, 153
\bibitem[Shalchi \& Gammon(2019)]{ShalchiEA2019}Shalchi, A, \& Gammon, M. 2019, Adv. Space Res., 63, 653 
\bibitem[Stawarz \& Petrosian(2008)]{StawarzEA2008}Stawarz, L., \& Petrosian, V. 2008, \apj, 681, 1725
\bibitem[Tautz \& Shalchi(2012)]{TautzEA2012}Tautz, R. C., \& Shalchi, A. 2012, \apj, 744, 125
\bibitem[Wang \& Qin(2018)]{wq2018}Wang, J.-F., Qin, G. 2018, \apj, 868, 139
\bibitem[Wang \& Qin(2019)]{wq2019}Wang, J.-F., Qin, G. 2019, \apj, 886, 89
\bibitem[Wang \& Qin(2020)]{wq2020}Wang, J.-F., Qin, G. 2020, \apj,  899, 39
\bibitem[Wang et al.(2017)]{WangEA2017}Wang, J.-F., Qin, G., Ma, Q.-M., Song, T., \& Yuan, S.-B. 2017, \apj, 845, 112
\bibitem[Wang \& Qin(2016)]{WangEA2016}Wang, Y., \& Qin, G. 2016, \apj, 820, 61
\bibitem[Zank et al.(2006)]{ZankEA2006}Zank, G. P., Li Gang, Florinski, V., Hu Qiang,
Lario, D., \& Smith, W. 2006, \jgr, 111, A06108
\bibitem[Zank et al. (2000)]{ZankEA2000}Zank, G. P., Rice, W. K. W, \& Wu, C. C. 2000,
\jgr, 105, 25079
\bibitem[Zhang (1999)]{ZhangM1999}Zhang M. 1999, \apj, 513, 409
\bibitem[Zimbardo et al.(2012)]{ZimbardoEA2012}Zimbardo, G., Perri, S., Pommois, P., 
\& Veltri, P. 2012, Advances in Space Research,  49, 1633 

\end{thebibliography}
\end{document}